\documentclass[12pt]{article}
\usepackage{comment}
\usepackage{amsmath}
\usepackage{amssymb}
\usepackage{graphicx}
\usepackage{here}
\usepackage{subcaption}
\usepackage{url}
\usepackage[sort&compress, numbers, merge]{natbib}
\usepackage{braket}
\usepackage{physics}
\usepackage[compat=1.1.0]{tikz-feynhand}
\usepackage{slashed}
\usepackage{mathtools}

\usepackage{todonotes}

\numberwithin{equation}{section}

\usepackage[colorlinks=true, linkcolor=blue, citecolor=blue,
urlcolor=black]{hyperref}

\begin{document}
\def\ps{\mathbf{p}}
\def\PS{\mathbf{P}}
\baselineskip 0.6cm
\def\simgt{\mathrel{\lower2.5pt\vbox{\lineskip=0pt\baselineskip=0pt
           \hbox{$>$}\hbox{$\sim$}}}}
\def\simlt{\mathrel{\lower2.5pt\vbox{\lineskip=0pt\baselineskip=0pt
           \hbox{$<$}\hbox{$\sim$}}}}
\def\simprop{\mathrel{\lower3.0pt\vbox{\lineskip=1.0pt\baselineskip=0pt
             \hbox{$\propto$}\hbox{$\sim$}}}}
\def\tr{\mathop{\rm tr}}
\def\SU{\mathop{\rm SU}}

\begin{titlepage}

\begin{flushright}
IPMU22-0024\\
KEK-TH-2420
\end{flushright}

\vskip 1.1cm

\begin{center}

{\Large \bf 
More on Fake GUT
}

\vskip 1.2cm
Masahiro Ibe$^{a,b}$, 
Satoshi Shirai$^{b}$,
Motoo Suzuki$^{c}$, 
Keiichi Watanabe$^{a}$ and
Tsutomu T. Yanagida$^{d}$
\vskip 0.5cm

{\it

$^a$ {ICRR, The University of Tokyo, Kashiwa, Chiba 277-8582, Japan}

$^b$ {Kavli Institute for the Physics and Mathematics of the Universe
 (WPI), \\The University of Tokyo Institutes for Advanced Study, \\ The
 University of Tokyo, Kashiwa 277-8583, Japan}
 
$^c$ {Institute of Particle and Nuclear Studies, High Energy Accelerator Research Organization (KEK), Tsukuba 305-0801, Japan}

$^d$ {Tsung-Dao Lee Institute and School of Physics and Astoronomy, \\Shanghai Jiao Tong University, 800 Dongchuan Road, Shanghai, 200240 China}
}

\vskip 1.0cm

\abstract{
It is remarkable that the matter fields in the Standard Model (SM) 
are apparently unified into the $\mathrm{SU}(5)$ representations.
A straightforward explanation of this fact is to embed all the SM gauge groups into a simple group containing $\mathrm{SU}(5)$, i.e., the grand unified theory (GUT).
Recently, however, a new framework ``fake GUT" has been proposed.   
In this new framework, 
the apparent matter unification can be explained by a chiral gauge group $G$, $ G \supset \mathrm{SU}(5)$.
We emphasize that the SM matter fields are not necessarily embedded into the chiral representations to explain the apparent unification. 
In this paper, we discuss details of
 concrete realizations of the fake GUT model.
We first study the model based on
$\mathrm{SU}(5) \times \mathrm{U}(2)_H$,
where $\mathrm{SU}(3)_c$ in the SM is from $\mathrm{SU}(5)$ while $\mathrm{SU}(2)_L\times \mathrm{U}(1)_Y$ are from the diagonal subgroups of $\mathrm{SU}(5) \times \mathrm{U}(2)_H$.
We also extend this model to the one based on a semi-simple group, $\mathrm{SU}(5) \times \mathrm{SU}(3)_H$, so that $\mathrm{U}(2)_H$
is embedded in $\mathrm{SU}(3)_H$.
We also show that this framework
predicts rather different decay patterns of the proton, compared to the conventional GUT.
}

\end{center}
\end{titlepage}

\section{Introduction} 

In the Standard Model (SM),
the matter fields are apparently unified into the $\mathrm{SU}(5)$ representations.
In general, chiral fermions consistent with SM gauge symmetry do not necessarily satisfy this property~\cite{Foot:1988qx,Knochel:2011ng,Cebola:2014qfa}.
Therefore, 
the apparent SM matter unification
into the $\mathrm{SU}(5)$ multiplets is quite remarkable.
In fact, the matter unification
has been propelling the study of the grand unified theory (GUT) for a long time \cite{Georgi:1974sy,Georgi:1974yf,Buras:1977yy} (see Ref.~\cite{ParticleDataGroup:2020ssz} for reviews).
In the conventional GUT, the SM matter fields are unified into common representations of a simple group containing $\mathrm{SU}(5)$.
As a result of the matter unification,
the GUT predicts the proton decay, which are extensively searched for in a variety of experiments (see e.g. Refs.\,\cite{Super-Kamiokande:2020wjk,Hyper-Kamiokande:2018ofw,JUNO:2015zny,DUNE:2020fgq}).

Recently, we have proposed a new framework ``fake GUT"~\cite{Ibe:2019ifm}.
This framework can explain the apparent unification of the SM matter fields into the $\mathrm{SU}(5)$ multiplets
in a different way than conventional GUT models.
In the fake GUT, the SM matter fields are not necessarily embedded into common $\mathrm{SU}(5)$ multiplets at the high energy.
Although the quarks and leptons can have different origins, they form complete $\mathrm{SU}(5)$ multiplets at the low energy as if they originate from the same multiplets.
In the fake GUT, the prediction of the proton decay can be significantly different from that in conventional GUT models.
As an extreme example, 
the quarks and leptons may reside in completely different multiplets.
In such a case, the proton decay does not occur.

In Ref.~\cite{Ibe:2019ifm}, 
a fake GUT model based on a non-simple $\mathrm{SU}(5) \times \mathrm{U}(2)_{H}$ group has been sketched.
In this model, the leptons mainly reside in the part of the vector-like fermions of $\mathrm{U}(2)_H$,
while the quarks reside in the $\overline{\mathbf{5}}\oplus \mathbf{10}$ representations.
After the spontaneous symmetry breaking of 
$\mathrm{SU}(5) \times \mathrm{U}(2)_{H}$
down to the SM gauge group, only the leptons and 
the quarks remain massless which apparently 
form $\overline{\mathbf{5}}\oplus \mathbf{10}$
representations.
Although the ``GUT scale" is predicted to be
much lower than the conventional GUT scale, the proton decay rate is suppressed as the quarks and the leptons are not in the same multiplets.
In this paper, we study details of the 
$\mathrm{SU}(5)\times \mathrm{U}(2)_H$ model.

We also extend the non-simple $\mathrm{SU}(5)\times \mathrm{U}(2)_H$ model to that based on a semi-simple group $\mathrm{SU}(5)\times\mathrm{SU}(3)_{H}$.
With this extension, we can successfully explain the charge quantization and avoid the Landau pole problem of $\mathrm{U}(1)_H$ gauge interaction.
We also discuss the symmetry which determines 
the fraction of the leptons originating from $\overline{\mathbf{5}}\oplus \mathbf{10}$, which 
in turn controls the proton decay rate.

The organization of this paper is as follows.
In Sec.\,\ref{sec:FAKE_GUT}, we first review the idea of the fake GUT.
In Sec.\,\ref{sec:U2_FAKE_GUT}, we discuss details of the non-simple $\mathrm{SU}(5)\times \mathrm{U}(2)_{H}$ group.
In Sec.\,\ref{sec:SU3_FAKE_GUT}, we extend the model to that based on a semi-simple $\mathrm{SU}(5)\times \mathrm{SU}(3)_{H}$ model.
The final section is devoted to our conclusions.

\section{Fake GUT}
\label{sec:FAKE_GUT}
In this section, we review the idea of the fake GUT.
Let us consider a high energy theory which satisfies the following conditions;
\begin{description}
    \setlength{\leftskip}{1.0cm}
    \item[$1$.]  The gauge group is $ G = \mathrm{SU}(5) \times H $, which is spontaneously broken down to the SM gauge group $G_{\mathrm{SM}} = \mathrm{SU}(3)_{c} \times \mathrm{SU}(2)_{L} \times \mathrm{U}(1)_{Y}$ at the fake GUT scale.
    \item[$2$.]  Three copies of the chiral fermions
    in the 
    $\overline{\mathbf{5}} \oplus \mathbf{10} $ representations of $\mathrm{SU}(5)$ 
    which are neutral under $H$.
    \item[$3$.] Additional fermions which consist of the vector-like representations of $G$.
    \item[$4$.] All the Cartan subgroups of $\mathrm{SU}(5)$ remain unbroken and all of them take 
    part in $G_{\mathrm{SM}}$.
    \item[$4'$.] Some of $\mathrm{SU}(3)_{c}$, $\mathrm{SU}(2)_{L}$ and $\mathrm{U}(1)_{Y}$ may be diagonal subgroups of $\mathrm{SU}(5) \times H $.
\end{description}
We call the theory which satisfies  conditions 1 to 4 the “fake GUT.”
In this work, we also assume 
the additional condition $4'$, which realizes more viable models
in view of the proton decay and 
the coupling unification.
As proven in Ref.\,\cite{Ibe:2019ifm},
the fake GUT model guarantees that 
the low energy fermions completely match with the $\overline{\mathbf{5}}\oplus \mathbf{10}$ representations, while the quarks and the leptons do not necessarily
originate from $\overline{\mathbf{5}}\oplus \mathbf{10}$ multiplets (see the App.~\ref{sec:Characters of SM group and SU(5)}).
When some of quarks/leptons originate from fields other than $\bar{\mathbf{5}}\oplus \mathbf{10}$,
some of the $\overline{\mathbf{5}}\oplus \mathbf{10}$ fermions become massive whose mass partners are parts of the vector-like fermions of $G$
after $G$ is spontaneously broken down to $G_\mathrm{SM}$.
In this way, only the SM fermions
remain massless at the GUT scale, which form the
$\overline{\mathbf{5}}\oplus \mathbf{10}$ representations
regardless of how they are embedded in the fake GUT representations at the high energy~\cite{Ibe:2019ifm}.
This feature strongly owes to the nature of the chiral fermion and does not apply to the bosonic fields.
In the case of supersymmetry (SUSY),
this argument applies to the chiral superfields.

Let us emphasize that
the apparent matter unification in the fake GUT is guaranteed  by 
the matching
of the  characters of the chiral representations and
does not 
depend on the details of the models. 
This argument is more versatile 
than the 't Hooft anomaly (i.e. the anomaly of the global symmetry) matching conditions 
to restrict the fermions in the low energy theory~\cite{tHooft:1979rat},
since the latter relies on the global symmetries of the model.

We may consider the fake GUT with a larger group $G_{\mathrm{UV}}$
than $\mathrm{SU}(5)\times H$,
as long as $G_{\mathrm{UV}}$ 
is broken to $\mathrm{SU}(5)\times H$ satisfying the above conditions.
For instance, $\mathrm{SO}(10)\times H \to \mathrm{SU}(5)\times H$ with the $\mathbf{16}$ representations 
satisfies the above conditions and works as a fake GUT model.

Another interesting feature of the fake GUT model is that it does not necessarily require the SM gauge coupling unification for a non-trivial $H$ when it satisfies the condition $4'$.
Thus, although the coupling unification fails in the non-supersymmetric SM,
the fake GUT framework works with a proper choice of $H$.

There are many possible 
choices of $H$.
For example, 
$H=1,\, \mathrm{U}(1)_H, \, \mathrm{SU}(N)_H \cdots$, 
are possible candidates.
It is also possible to consider 
$H = \mathrm{U}(2)_{H},\,\mathrm{U}(3)_{H},\,\mathrm{SU(3)}_H\times \mathrm{SU(2)}_H \times \mathrm{U}(1)_H$,
where $\mathrm{U}(2)_{H}\supset \mathrm{SU}(2)_L\times \mathrm{U}(1)_Y$,  $\mathrm{U}(3)_{H}\supset \mathrm{SU}(3)_c\times \mathrm{U}(1)_Y$ etc.
In fact, many extensions of the GUT models fit into the fake GUT framework.
For instance, massive fermion extensions play important role in the Yukawa coupling unification~\cite{Murayama:1991ew,Hisano:1993uk}, (see also Refs.~\cite{Chigusa:2017drd,Hall:2018let,Hall:2019qwx} for recent applications), 
and in suppressing lepton/baryon number violation
in GUT~\cite{Bhattacherjee:2013gr,Fornal:2017xcj,Cacciapaglia:2020qky}.

Not all possibilities are, however, phenomenologically viable due to the constraints from the gauge coupling matching conditions and the proton lifetime.
As we will see later, we find that the minimal viable choice is 
$H = \mathrm{U}(2)_{H}$,
if the running of the gauge coupling constants is the same as the SM below the fake GUT scale.
In this case, the smaller choices 
such as
$H=1$, $H=\mathrm{U}(1)_H$ or $H=\mathrm{SU}(2)_H$ are not phenomenologically compatible. 
In the next section, we discuss a $\mathrm{SU}(5)\times \mathrm{U}(2)_H$ model.
We also extend the model so that $\mathrm{U}(2)_H$ is embedded into $\mathrm{SU}(3)_H$ in Sec.\,\ref{sec:SU3_FAKE_GUT}.

\section{\texorpdfstring{$\mathrm{SU}(5) \times \mathrm{U}(2)_{H}$}{} model}
\label{sec:U2_FAKE_GUT}

\begin{table}[t!]
\caption{The content of the fermions, the scalar fields and the gauge bosons in the $\mathrm{SU}(5) \times \mathrm{U}(2)_{H}$ model is shown in the group representation, $(\mathrm{SU}(5), \mathrm{SU}(2)_{H})_{\mathrm{U}(1)_{H}}$ and $(\mathrm{SU}(3)_c, \mathrm{SU}(2)_{L})_{\mathrm{U}(1)_{Y}}$.
Each fermion has three generations.
The $\mathrm{U}(1)_Y$ neutral scalar fields and gauge bosons are given by the real scalar fields and real gauge bosons. 
The scalar fields enclosed in square brackets 
are the Goldstone modes associated with 
$\mathrm{SU}(5)\times \mathrm{U}(2)_H$ breaking 
to $G_\mathrm{SM}$, which 
are eaten by the gauge bosons.}
  \label{tab:U2particlecontents}
 \begin{center}
  \begin{tabular}{|c||c|c|}  \hline
     & $(\mathrm{SU}(5),\mathrm{SU}(2)_{H})_{\mathrm{U}(1)_{H}}$ & $(\mathrm{SU}(3)_{c},\mathrm{SU}(2)_{L})_{\mathrm{U}(1)_{Y}}$ \\ \hline \hline
    fermions & &  \\ \hline 
    $\overline{\mathbf{5}}$ & $(\overline{\mathbf{5}},\bf{1})_{\mathrm{0}}$ & $(\mathbf{\overline{3}},\bf{1})_{\mathrm{+1/3}} \oplus (\mathbf{1},\mathbf{2})_{\mathrm{-1/2}}$ \\ 
    $\mathbf{10}$ & $(\mathbf{10},\bf{1})_{\mathrm{0}}$ & $(\mathbf{3},\mathbf{2})_{\mathrm{+1/6}} \oplus (\mathbf{\overline{3}},\mathbf{1})_{\mathrm{-2/3}} \oplus (\mathbf{1},\mathbf{1})_{\mathrm{+1}}$ \\
    $L_{H}$ & $(\mathbf{1},\mathbf{2})_{-1/2}$ & $(\mathbf{1},\mathbf{2})_{\mathrm{-1/2}}$ \\
    $\overline{L}_{H}$ & $(\mathbf{1},\mathbf{2})_{+1/2}$ & $(\mathbf{1},\mathbf{2})_{\mathrm{+1/2}}$ \\ 
    $E_{H}$ & $(\mathbf{1},\mathbf{1})_{-1}$ & $(\mathbf{1},\mathbf{1})_{\mathrm{-1}}$  \\
    $\overline{E}_{H}$ & $(\mathbf{1},\mathbf{1})_{+1}$ & $(\mathbf{1},\mathbf{1})_{\mathrm{+1}}$ \\ \hline \hline
    scalars & &  \\ \hline
    $\phi_{2}$ & $(\mathbf{5}, \mathbf{2})_{-1/2}$ & $  (\mathbf{1},\mathbf{3})_{\mathrm{0}}
    \oplus (\mathbf{1},\mathbf{1})_{\mathrm{0}}
     \oplus [
      (\mathbf{3},\mathbf{2})_{\mathrm{-5/6}}
      \oplus
      (\mathbf{1},\mathbf{3})_{\mathrm{0}}
     \oplus (\mathbf{1},\mathbf{1})_{\mathrm{0}} ]
    $ \\
    $H_{5}$ & $(\mathbf{5}, \mathbf{1})_{0}$ & $(\mathbf{3},\mathbf{1})_{\mathrm{-1/3}} \oplus (\mathbf{1},\mathbf{2})_{\mathrm{+1/2}}$ \\
    $H_{2}$ & $(\mathbf{1}, \mathbf{2})_{+1/2}$ & $(\mathbf{1},\mathbf{2})_{\mathrm{+1/2}}$ \\ \hline \hline
    vectors & &  \\ \hline
    $V_{5}$ & $(\mathbf{24}, \mathbf{1})_{0}$ & $(\mathbf{8},\mathbf{1})_{\mathrm{0}} \oplus (\mathbf{3},\mathbf{2})_{\mathrm{-5/6}} \oplus 
    (\mathbf{1},\mathbf{3})_{\mathrm{0}} \oplus (\mathbf{1},\mathbf{1})_{\mathrm{0}}$ \\
    $V_{2H}$ & $(\mathbf{1}, \mathbf{3})_{0}$ & $(\mathbf{1},\mathbf{3})_{\mathrm{0}}$ \\
    $V_{1H}$ & $(\mathbf{1}, \mathbf{1})_{0}$ & $(\mathbf{1},\mathbf{1})_{\mathrm{0}}$ \\ \hline
  \end{tabular}
 \end{center} 
\end{table}

In this section,
we discuss a model with $G = \mathrm{SU}(5) \times \mathrm{U}(2)_{H}$, where 
$\mathrm{SU}(5) \supset
\mathrm{SU}(3)_c\times 
\mathrm{SU}(2)_L\times
\mathrm{U}(1)_Y
$
and 
$\mathrm{U}(2)_H \supset\mathrm{SU}(2)_L \times \mathrm{U}(1)_Y$.
Below the fake GUT scale, $\mathrm{SU}(2)_L \times \mathrm{U}(1)_Y$ appear as the diagonal subgroups 
of $\mathrm{SU}(5) \times \mathrm{U}(2)_{H}$, while $\mathrm{SU}(3)_c$ appears solely from $\mathrm{SU}(5)$.
As we emphasized above, this choice is the minimal gauge group for the
fake GUT model which 
is phenomenologically compatible.
We summarize the matter contents of this model in Tab.~\ref{tab:U2particlecontents}.

Note that the idea of the fake GUT is completely different from
that of the SUSY GUT model based on the product group~\cite{Yanagida:1994vq},
although they are both based on the product gauge group.
In the product group SUSY GUT model, 
the product group is used to solve 
the so-called the doublet-triplet mass splitting problem, while the effective gauge coupling unification is taken seriously and the matter multiplets are assumed as the GUT multiplets.

\subsection{Origin of SM fermions}
We introduce the three generations of the chiral multiplets $\overline{\mathbf{5}}\oplus \mathbf{10}$ of $\mathrm{SU}(5)$.
The SM right-handed down quarks $\overline{d}_R$ sector fully come from the $\overline{\mathbf{5}}$,
and the right-handed up quarks $\overline{u}_R$ and the left-handed $\mathrm{SU}(2)_L$ doublet quarks $q_L$ from $\mathbf{10}$.%
\footnote{In this paper, we use Weyl fermion notation thoroughly.}
In addition,
we introduce three pairs of vector-like multiplets charged under $H =$ U$(2)_{H}$,
\begin{gather}
\label{eq:U2 lepton doublet}
(L_{H}:(\mathbf{1},\mathbf{2})_{-1/2},\,\,\,\,\, \overline{L}_{H}:(\mathbf{1},\mathbf{2})_{+1/2}) \,\times\, 3, \\
\label{eq:U2 lepton singlet}
(E_{H}:(\mathbf{1},\mathbf{1})_{-1},\,\,\,\,\, \overline{E}_{H}:(\mathbf{1},\mathbf{1})_{+1})\,\times\, 3.
\end{gather}
Here, the group representations are denoted by $(\mathrm{SU}(5), \mathrm{SU}(2)_{H})_{\mathrm{U}(1)_{H}}$.

In this model, spontaneous breaking
of SU$(5)\times$U$(2)_H$ into $G_{\mathrm{SM}}$
is achieved by a vacuum expectation value (VEV) of a complex scalar field $\phi_{2}$,
which is a bi-fundamental representation, 
$(\mathbf{5}, \mathbf{2})_{-1/2}$.
Explicitly, the VEV
of $\phi_{2}$ (see Ref.~\cite{Hotta:1995cd})
\begin{align}
\label{eq: Phi vev} 
\ev{\phi_{2}} = 
 \left( \begin{array}{ccccc}
      0 & 0 & 0 & v_{2} & 0 \\
      0 & 0 & 0 & 0 & v_{2}
    \end{array} \right),
\end{align}
breaks SU$(5)\times$U$(2)_H$ into $G_{\mathrm{SM}}$.
Here, $v_{2} > 0$ is a constant with mass dimension and much larger than the electroweak scale.
In this case, 
$\mathrm{SU}(3)_{c}$ appears as an unbroken subgroup of $\mathrm{SU}(5)$,
while $\mathrm{SU}(2)_{L}$ and $\mathrm{U}(1)_{Y}$ appear as diagonal subgroups of $\mathrm{SU}(5)$ and $\mathrm{U}(2)_{H}$.
After the symmetry breaking, 
the gauge bosons corresponding to the broken 
generators obtain masses of $\mathcal{O}(v_2)$.
The gauge charges and the masses of them are
given in Tab.\,\ref{tab:U2vectormass}.

\begin{table}[tb]
\caption{The mass spectrum of the gauge bosons.
$X_{\mu}$, $\Omega_{3\mu}$ and $\Omega_{1\mu}$ are the $\mathrm{SU}(5)$ gauge fields, $\mathrm{SU}(2)_{H}$ gauge fields, and $\mathrm{U}(1)_{H}$ gauge fields respectively.}
  \label{tab:U2vectormass}
 \begin{center}
  \begin{tabular}{|c||c|c|}  \hline
    vectors & $(\mathrm{SU}(3)_{c},\mathrm{SU}(2)_{L})_{\mathrm{U}(1)_{Y}}$ & mass \\ \hline \hline
    $X_{\mu}$ & $(\mathbf{3},\mathbf{2})_{\mathrm{-5/6}}$ & $g_{5}v_{2}/\sqrt{2}$ \\
    $\Omega_{3\mu}$ & $(\mathbf{1}, \mathbf{3})_{0}$ & $\sqrt{g_{5}^{2} + g_{2H}^{2}} v_{2}$ \\
    $\Omega_{1\mu}$ & $(\mathbf{1}, \mathbf{1})_{0}$ & $\sqrt{3g_{5}^{2}/5 + g_{1H}^{2}} v_{2}$ \\ \hline
  \end{tabular}
 \end{center} 
\end{table}
Once $\mathrm{SU}(5) \times \mathrm{U}(2)_{H}$ is broken, 
the massless fermions coincide to the SM chiral fermions, 
while the other fermions become heavy.
To see this point explicitly, 
let us consider the following interactions 
between the fermions and $\phi_{2}$,
\begin{align}
\label{eq:Abelian mass}
\mathcal{L} 
&= m_{L,ij} \overline{L}_{Hi} L_{Hj} + \lambda_{L,ij} \overline{L}_{Hi} \phi_{2} \overline{\mathbf{5}}_{j} 
+ m_{E,ij} E_{Hi} \overline{E}_{Hj} + \frac{\lambda_{E,ij}}{\Lambda_{\mathrm{cut}}} E_{Hi} \phi_{2}^{\dagger} \phi_{2}^{\dagger} \mathbf{10}_j + h.c..
\end{align}
Here, $\lambda_{L,E}$ are coupling constants, and
$\Lambda_{\mathrm{cut}}$ a cutoff scale
larger than the fake GUT scale.
The summation of the flavor indices $i,j=1,2,3$ is understood.
The above interactions are the most general forms of fermion bilinears up to mass dimension five.
The higher dimensional operator can be generated by integrating over the massive 
fermions in $(\mathbf{5},\mathbf{2})_{-1/2}$ representation coupling to $\phi_2$.
Or, we may also consider a complex scalar
$\phi_2'$ in $(\overline{\mathbf{10}},\mathbf{1})_{1}$ 
which has a Yukawa coupling ${E}_H\phi_2' \mathbf{10}$
and a trilinear coupling $\phi_2 \phi_2' \phi_2$. 

After the fake GUT symmetry breaking,
the above interactions lead to 
the mass terms in the leptonic sector. The quarks are, on the other hand, fully contained in $\overline{\mathbf{5}}\oplus\mathbf{10}$, and hence, remain massless.
The mass terms of the leptonic sector are given by,
\begin{align}
\label{eq: mixing}
  \mathcal{L}_{\mathrm{mass}}
  = \overline{L}_{Hi}
  \,\mathcal{M}_{L,ij}
   \left(
    \begin{array}{c}
      \overline{\mathbf{5}}_{L} \\
      L_{H}
    \end{array}
  \right)_j + 
  E_{Hi} \,\mathcal{M}_{E,ij}
  \left(
    \begin{array}{c}
      \mathbf{10}_{\overline{E}} \\
      \overline{E}_{H}
    \end{array}
  \right)_{j} +
  h.c.,
  \end{align}
  where the mass matrices are given written as
 \begin{align}
 \label{eq:lepton_mass}
     \mathcal{M}_{L,ij} = 
        \left(
    \begin{array}{cc}
      \lambda_{L,ij} v_{2} & m_{L,ij}
    \end{array}
  \right), \quad 
     \mathcal{M}_{E,ij} = 
         \left(
     \begin{array}{cc}
       \displaystyle{\frac{\lambda_{E,ij} v_{2}^{2}}{\Lambda_{\mathrm{cut}}}} & m_{E,ij}
     \end{array}
    \right) \ .
 \end{align}
Here,
$\overline{\mathbf{5}}_{L}$ and 
$\mathbf{10}_{\overline{E}}$ 
denote the components of 
the $G_{\mathrm{SM}}$ gauge charges corresponding to those
of the doublet and the singlet leptons.
Since the mass matrices are given by  $3\times 6$ matrices, we find that three leptons remain massless due to the rank conditions.
As a result, the three generations of the Weyl fermions of the SM can be realized, although the leptons do not fully belong to 
$\overline{\mathbf{5}}\oplus\mathbf{10}$.

To see the origin of the massless leptons, let us omit the flavor mixing and focus on a single generation.
In this case, the mass eigenstates are given by 
\begin{align}
\label{eq:L mixing}
  &\left(
    \begin{array}{c}
      L_{M} \\
      \ell_{L}
    \end{array}
  \right) 
  = \left(
    \begin{array}{cc}
      \mathrm{cos}\, \theta_{L} & \mathrm{sin}\, \theta_{L} \\
      -\mathrm{sin}\, \theta_{L} & \mathrm{cos}\, \theta_{L}
    \end{array}
  \right) 
  \left(
    \begin{array}{c}
      \overline{\mathbf{5}}_{L} \\
      L_{H}
    \end{array}
  \right), \\
  \label{eq:E mixing}
  &\left(
    \begin{array}{c}
      \overline{E}_{M} \\
      \overline{e}_{R}
    \end{array}
  \right) 
  = \left(
    \begin{array}{cc}
      \mathrm{cos}\, \theta_{E} & \mathrm{sin}\, \theta_{E} \\
      -\mathrm{sin}\, \theta_{E} & \mathrm{cos}\, \theta_{E}
    \end{array}
  \right) 
  \left(
    \begin{array}{c}
      \mathbf{10}_{\overline{E}} \\
      \overline{E}_{H}
    \end{array}
  \right) ,
  \end{align}
where $\ell_L$ and $\overline{e}_R$ are the doublet and the singlet massless eigenstates, respectively.
The mixing angles, $\theta_{L,E}$, are
\begin{align}
\label{eq:L_mixing_angle}
\mathrm{tan}\, \theta_{L} 
&= \frac{ m_{L}}{\lambda_{L} v_{2}}, \\
\label{eq:E_mixing_angle}
\mathrm{tan}\, \theta_{E} 
&= \frac{ m_{E} \, \Lambda_{\mathrm{cut}}}{\lambda_{E} v_{2}^{2}},
\end{align}
where we have taken the parameters real positive.
The masses of the heavy leptons are given by,
\begin{align}
    M_L = \sqrt{\lambda_L^2 v_{2}^2 + m_L^2}\ , \quad 
     M_E = \sqrt{\lambda_E^2 v_{2}^4/\Lambda_{\mathrm{cut}}^{2} + m_E^2 }\ .
\end{align}

As an extreme example, the SM leptons are fully contained in 
$L_H$ and $\overline{E}_H$ for $m_L = m_E = 0$, while the heavy leptons remain massive.
In this case, although the quarks and the leptons in the SM originate from 
completely separate multiplets in the fake GUT, the quarks and the leptons apparently form $\overline{\mathbf{5}}\oplus \mathbf{10}$ multiplets at the low energy.
Note that the limit $m_L=m_E=0$ 
enhances a global symmetry under which $L_H$ and $\overline{E}_H$ are charged (see Sec.\,\ref{sec:Global Symmetry}).

\subsection{Origin of SM Higgs and Yukawa interactions}
\label{sec:Yukawa interaction}

In the previous subsection, we discussed the origin of the SM Weyl fermions.
Here, 
we discuss the origin of the SM Higgs boson and the Yukawa interactions.
As the quarks and the leptons can originate from the separate multiplets,
the SM Yukawa interactions consist of various contributions.
Here, we consider a case that only one SM Higgs doublet remains in the low energy.

Concretely, we introduce $H_{5}$ and $H_{2}$ scalar fields,
which are of 
$(\mathbf{5}, \mathbf{1})_{0}$ and $(\mathbf{1}, \mathbf{2})_{1/2}$ 
representations.
These scalar fields contain the SM Higgs scalars as well as the colored Higgs,
\begin{align}
    H_2 = h_2^{\mathrm{SM}} \ , \quad H_5 = 
    \left(
\begin{array}{c}
     h_5^{\mathrm{color}}  \\
     h_5^{\mathrm{SM}} 
\end{array}
    \right)\ .
\end{align}
The SM Higgs components in $H_5$ and $H_2$ mix with each other, via an interaction,
\begin{align}
\label{eq:Higgs_mixing}
\mathcal{L}_{52 \mathrm{mix}} 
= \mu_{\mathrm{mix}} H_{2} \phi_{2} {H}_{5}^{*} + h.c.,
\end{align}
once $\phi_2$ develops the VEV.
As a result, 
the effective mass terms of $H_5$ and $H_2$ are given by, 
\begin{align}
    \mathcal{L} = - m_5'^{2} |h_5^{\mathrm{color}}|^2 - m_5^2 |h_5^{\mathrm{SM}}|^2 - m_2^2 |h_2^{\mathrm{SM}}|^2 + (\mu_{\mathrm{mix}} v_{2} h_2^{\mathrm{SM}} h_5^\mathrm{SM*} + h.c.) \ .
\end{align}
Accordingly, the SM Higgs, $h^{\rm SM}$, is given by a linear combination of $h_2^{\mathrm{SM}}$ and $h_5^{\mathrm{SM}}$,
\begin{align}
\label{eq:higgs_mixing}
h^{\rm SM}=\cos\theta_h\, h_2^{\rm SM}-\sin\theta_h\, h_5^{\rm SM}\ .
\end{align}
Here, $m_5^2$, $m_5'^2$, and $m_2^2$
are orders of $\order{v_{2}^2}$ and 
include the contribution from the VEV of $\phi_{2}$.
We also assume $\mu_{\mathrm{mix}}$ is of $\order{v_{2}}$.
To achieve the mass of 
$h^{\mathrm{SM}}$ in $\order{100}$\,GeV, 
we require severe fine-tuning
as in the case of the conventional GUT.

By using $H_5$ and $H_2$, 
the origins of the SM Yukawa interactions are given by,
\begin{align}
\label{eq:quark Yukawa}
\mathcal{L}_{YQ} 
&= -(y_{5})_{ij}\, \overline{\mathbf{5}}_{i}\, \mathbf{10}_{j}\, {H}_{5}^{*}
- (y_{10})_{ij}\, \mathbf{10}_{i}\, \mathbf{10}_{j}\, {H}_{5}
+ h.c.\ , \\
\label{eq:lepton Yukawa}
\mathcal{L}_{YL} 
&= -(y_{LE})_{ij} \, L_{Hi}\, \overline{E}_{Hj}\, H_{2}^{*}
+ h.c.\ ,
\end{align}
where $i$ and $j$ run the number of generations.
The SM Yukawa couplings are obtained by 
substituting $h_5^\mathrm{SM} \to - \sin\theta_h h^{\mathrm{SM}}$, $h_2^\mathrm{SM} \to \cos\theta_h h^{\mathrm{SM}}$ after diagonalizing the mass matrices in Eq.\,\eqref{eq:lepton_mass}.

As we will see in Sec.\,\ref{sec:Nucleon decay},
the lepton components in $\overline{\mathbf{5}}\oplus \mathbf{10}$ should be highly suppressed 
to evade the constrains from the proton decay,
i.e., $\theta_{L,E}\ll 1$.
In this case, the SM Yukawa couplings are approximately given by, 
\begin{align}
\label{eq:Yukawa Origins}
\begin{split}
(y^{\mathrm {SM}}_u)_{ij} =& -\sin\theta_h (y_{10})_{ij}\ ,\\
(y^{\mathrm {SM}}_d)_{ij} =& -\sin\theta_h (y_{5})_{ij}\ ,\\
    (y^{\mathrm {SM}}_e)_{ij} =& \cos\theta_h (y_{LE})_{ij} + \order{\theta_{L}\theta_{E}} \sin\theta_h(y_5)_{ij}\ .
    \end{split}
\end{align}

\subsection{Gauge coupling constants}
\label{sec:gauge coupling in Abelian FAKE GUT}
\begin{figure}[t!]
	\centering
	\includegraphics[width=0.6\textwidth]{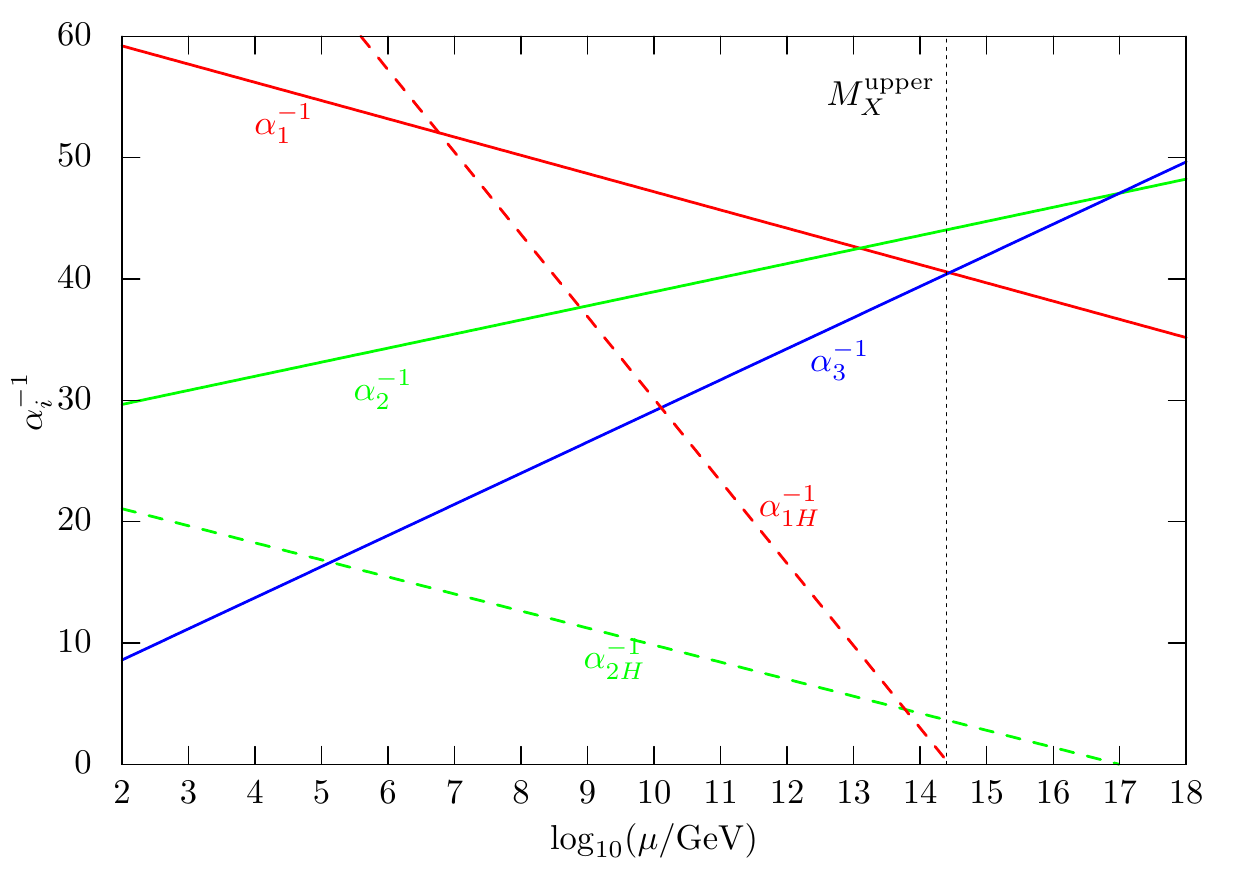}
\caption{The running of the SM gauge couplings (solid lines).
The red and green dashed lines denote the matched values of $\alpha_{2H}$ and $\alpha_{1H}$ for a given fake GUT scale (see Eqs.\,\eqref{eq:matching}).
The fake GUT scale is required to be below $M_X^\mathrm{upper} \simeq 10^{14.4}$\,GeV.
}
\label{fig:SM_running}
\end{figure}

In the present setup,
$\mathrm{SU}(2)_{L} \times \mathrm{U}(1)_{Y}$ gauge symmetries are the diagonal subgroups of $\mathrm{SU}(5) \times \mathrm{U}(2)_{H}$, and $\mathrm{SU}(3)_{c}$ gauge symmetry is the remaining unbroken subgroup of $\mathrm{SU}(5)$.
The tree-level matching conditions of the gauge coupling constants of the SM and the $\mathrm{SU}(5) \times \mathrm{U}(2)_{H}$ model at the fake GUT scale 
are given by~\cite{Ibe:2003ys},
\begin{align}
\label{eq:matching}
\begin{split}
\alpha_{1}^{-1}(M_{X}) &= \alpha_{5}^{-1}(M_{X}) + \frac{3}{5} \alpha_{1H}^{-1}(M_{X}), \\
\alpha_{2}^{-1}(M_{X}) &= \alpha_{5}^{-1}(M_{X}) + \alpha_{2H}^{-1}(M_{X}), \\
\alpha_{3}^{-1}(M_{X}) &= \alpha_{5}^{-1}(M_{X}),
\end{split}
\end{align}
where $\alpha_{i} = g_{i}^{2} / (4\pi)$, and $g_{i}$'s are the gauge coupling constants.
Here, $M_{X}$ is the mass of $X_\mu$ in Tab.\,\ref{tab:U2vectormass}, and we assume that
all the supermassive particles have the masses
of the same order of the magnitude.
The fine structure constants $\alpha_{5}, \alpha_{2H}$ and $\alpha_{1H}$ are those of $\mathrm{SU}(5)$ and $\mathrm{SU}(2)_H\times \mathrm{U}(1)_H$, respectively.%
\footnote{Here, $\alpha_{1}$ satisfies $\alpha_{1} = 5/3 \, \alpha_{Y}$, where $\alpha_{Y}$ is the fine structure constant of $\mathrm{U}(1)_{Y}$.}
Thus, the gauge couplings in the SM do not unify at the fake GUT scale.

As an interesting feature of $\mathrm{SU}(5)\times \mathrm{U}(2)_H$ model, there is an upper-limit on the fake GUT scale.
From the matching conditions in Eq.\,\eqref{eq:matching}, the fake GUT scale is consistent only for $\alpha_1^{-1} > \alpha_3^{-1}$ 
and $\alpha_2^{-1} > \alpha_3^{-1}$, 
since otherwise either $\alpha_{1H}$ or $\alpha_{2H}$ is negative.
From Fig.\,\ref{fig:SM_running}, we find
the fake GUT scale is lower than $M_X^{\rm upper}\simeq 10^{14.4}$\,GeV at which $\alpha^{-1}_1=\alpha^{-1}_3$.
On the other hand, there is no lower limit on the fake GUT scale, $M_X$, from the coupling matching condition.
The phenomenological lower limits on $M_X$ are
of $\order{10^{4\mbox{--}5}}$\,GeV from
collider experiments and precision measurements.

Here, let us comment on the coupling matching condition for other choice of the gauge group $H$.
For the model with $\mathrm{SU}(5)\times\mathrm{U}(3)_H$
where $\mathrm{SU}(3)_c \times \mathrm{U}(1)_{Y} \subset \mathrm{U}(3)_H$,
for example, the matching condition of the gauge couplings similar to Eq.\,\eqref{eq:matching}, requires that $\alpha_{1,3}^{-1} > \alpha_2^{-1}$.
This is not possible unless $G_{\mathrm{SM}}$ charged particles other than the SM fields appear below the fake GUT scale (see Fig.\,\ref{fig:SM_running}).
In this way, the matching condition
and the gauge coupling running of the SM 
constraints the choice of $H$.

\subsection{Global lepton and baryon symmetries}
\label{sec:Global Symmetry}

\begin{table}[t!]
\caption{The global $\mathrm{U}(1)_{5}$ and $\mathrm{U}(1)_{LH}$ charges of fermions and scalars are shown.
Here, $\mathrm{U}(1)_{LH}$ charges of $m_{L,E}$ in  Eq.~\eqref{eq:Abelian mass} are charges of spurions. }
  \label{tab:global_symmetry}
 \begin{center}
  \begin{tabular}{|c||c|c|c|c|c|c||c|c|c||c|c|c|}  \hline
 & $\overline{\mathbf{5}}$ & $\mathbf{10}$ & $L_{H}$ & $\overline{L}_{H}$ & $E_{H}$ & $\overline{E}_{H}$ & $\phi_{2}$ & $H_{5}$ & $H_{2}$ & $m_{L}$ & $m_{E}$ \\ \hline \hline
    $\mathrm{U}(1)_{5}$ & $-3$ & $1$ & $-3$ & $3$ & $-1$ & $1$ & $0$ & $-2$ & $-2$ & $0$ & $0$ \\ \hline 
    $\mathrm{U}(1)_{LH}$ & $0$ & $0$ & $1$ & $0$ & $0$ & $-1$ & $0$ & $0$ & $0$ & $-1$ & $1$\\ \hline 
  \end{tabular}
 \end{center} 
\end{table}
The present model possesses
the global U(1)$_5$ symmetry (fiveness) 
as in the conventional GUT model.
The charge assignment of the fiveness enlarged to the $\mathrm{U}(2)_H$ sector is given in Tab.\,\ref{tab:global_symmetry}.
Note that the interaction terms
in Eqs.~\eqref{eq:Abelian mass}, \eqref{eq:Higgs_mixing}, 
\eqref{eq:quark Yukawa},
and \eqref{eq:lepton Yukawa} allow two independent $\mathrm{U}(1)$ symmetries one of which is gauged as $\mathrm{U}(1)_H$ symmetry. 
Thus, the uniqueness of charge assignment of $\mathrm{U}(1)_5$ is up to the $\mathrm{U}(1)_H$ charge. We fix the $\mathrm{U}(1)_5$ charge by taking the $\mathrm{U}(1)_5$ charge of $\phi_2$ vanishing.

After the $\mathrm{SU}(5)\times \mathrm{U}(2)_H$
breaking,
$\mathrm{U}(1)_5$ results in 
the low energy $\mathrm{U}(1)_{B-L}$ symmetry,
with the charge 
\begin{align}
 Q_{B-L} = \frac{1}{5}(Q_5 + 4 Q_Y) \ ,
\end{align}
where $Q_5$ and $Q_Y$ are the charges of $\mathrm{U}(1)_5$ and $\mathrm{U}(1)_Y$.
The $B-L$ applies to the fermion fields in the same way of the SM.

As mentioned earlier, 
an additional global symmetry is enhanced
in the limit of $m_{L,E}\to 0$,
which we call $\mathrm{U}(1)_{LH}$
symmetry. 
The charge assignment is given in Tab.\,\ref{tab:global_symmetry}.
The $\mathrm{U}(1)_{LH}$ symmetry
also remains in the low energy which gives
an additional lepton symmetry, $\mathrm{U}(1)_L$.
This lepton symmetry is, however,
not exact since it 
is broken by the gauge/gravitational anomalies.
This additional symmetry can be identified with the lepton symmetry of the SM.

Up to the gauge/gravitational anomalies
the $\mathrm{U}(1)_{B-L}$ 
and $\mathrm{U}(1)_{L}$ symmetries are conserved separately in the limit of $m_{L,E}\to 0$, which may be rearranged as $\mathrm{U}(1)_B$ and $\mathrm{U}(1)_L$.
As far as these symmetries are respected, no visible low energy baryon/lepton violating processes are expected.
Note that the effects of the breaking of $\mathrm{U}(1)_{B-L}$ and $\mathrm{U}(1)_{L}$ by the gauge and the gravitational anomalies are highly suppressed for the nucleus decays~\cite{tHooft:1976rip}.

In the conventional $\mathrm{SU}(5)$ GUT
both the baryon symmetry $\mathrm{U}(1)_B$ and the lepton symmetry $\mathrm{U}(1)_{L}$ are embedded in the GUT fiveness.
Thus, at the low energy, only the linear combination, $\mathrm{U}(1)_{B-L}$, 
is conserved.
As a result, the conventional GUT predicts the proton decay which
violates the $\mathrm{U}(1)_B$ and $\mathrm{U}(1)_{L}$ symmetries
while $\mathrm{U}(1)_{B-L}$ is conserved.
The proton lifetime is dominantly determined by the GUT gauge boson mass scale.

In the present fake GUT model, the proton decay is forbidden 
when the $\mathrm{U}(1)_{LH}$ symmetry  emerges, i.e. $m_{L,E}=0$.
In fact, this symmetry is crucial for the successful $\mathrm{U}(2)_H$ fake GUT model as the 
$X$ gauge boson mass scale $M_X < 10^{14.4}\,$GeV, which would lead to too rapid proton decay without this symmetry.

It is, however, highly non-trivial whether we can impose such global symmetries on the model.
In general, it is argued that all global symmetries are broken by quantum gravity effects\,(see e.g. \cite{Hawking:1987mz,Lavrelashvili:1987jg,Giddings:1988cx,Coleman:1988tj,Gilbert:1989nq,Kallosh:1995hi,Banks:2010zn}).
Moreover, the gauge/gravitational 
anomalies  already break the $\mathrm{U}(1)_{B-L}$ and $\mathrm{U}(1)_{L}$ symmetries explicitly
in this model.
In addition, the origins of the neutrino masses and the baryon asymmetry in the Universe may also indicate the breaking of those symmetries.
Therefore, we expect that small violation of $\mathrm{U}(1)_{B-L}$ and $\mathrm{U}(1)_{L}$ exist which generate tiny $m_{L,E}$.
The breaking of the $\mathrm{U}(1)_{B-L}$ and $\mathrm{U}(1)_{L}$ symmetries depend on the further high energy physics.
In the following analysis, we simply regard $m_{L,E}$ (or equivalently $\theta_{L,E}$ in Eqs.\,\eqref{eq:L_mixing_angle} and \eqref{eq:E_mixing_angle}) 
as the parameter of the explicit symmetry breaking.%
\footnote{We discuss one example of the origin of the $\mathrm{U}(1)_{LH}$ symmetry and its breaking based on the $\mathrm{SU}(5)\times \mathrm{SU}(3)_H$ model
in the App.~\ref{sec:On global lepton and baryon symmetry}.}

Let us also comment on the origin of the neutrino mass.
The observations of the neutrino oscillations show that the neutrinos have tiny masses.
In the present model, there are various ways to realize the active neutrino masses.
First, we may consider the active neutrinos as the Dirac neutrinos,
where the right-handed neutrinos couple to 
the leptons and Higgs in the $\mathrm{U}(2)_H$ sector.
In this case, there is no lepton symmetry violation, and hence, the proton decay is also suppressed for 
$m_{L,E} \to 0$ as explained above.

Another possibility is to couple the right-handed neutrinos to the $\mathrm{SU}(5)$ sector.
In this case, although the active neutrino masses are the Dirac type,
we require the lepton symmetry breaking,
that is,
the tiny mixing between the 
lepton components of $\overline{\mathbf{5}}$
and $L_H$.
It is an interesting feature of this model,
that both the neutrino masses and 
the proton decay rate are generated 
by the effects of the lepton symmetry breaking, i.e., $m_{L}$.

More attractive possibility is 
to consider the Majorana neutrino masses,
which break $\mathrm{U}(1)_{LH}$ symmetry
down to $\mathbb{Z}_2$ subgroup.
When $\mathbb{Z}_2$ subgroup of $\mathrm{U}(1)_{LH}$ is conserved separately from  $\mathrm{U}(1)_{B-L}$, 
the proton is stable
as long as $Z_2$ is unbroken.
For instance, 
it is possible to consider 
the seesaw mechanism \cite{Minkowski:1977sc,Yanagida:1979as,*Yanagida:1979gs,Gell-Mann:1979vob,Glashow:1979nm,Mohapatra:1979ia} in the $\mathrm{U}(2)_H$ sector where 
the model possesses the $Z_2$ symmetry 
while $m_{L,E}$ are suppressed.

\subsection{Nucleon decay}
\label{sec:Nucleon decay}
The nucleons can decay through the heavy $\mathrm{SU}(5)$ gauge bosons $X$
exchange.
For simplicity,
we assume $\theta_{E,L}$ in Eq.\,\eqref{eq:L mixing} and \eqref{eq:E mixing} do not depend on the generations.
In this case,
the proton lifetime of the $ p \rightarrow \pi^{0} + e^{+}$ mode is 
\begin{align}
\label{eq:proton_lifetime_in_U2 }
\tau(p \rightarrow \pi^{0} + e^{+}) \simeq 
 \frac{5\times 10^{26} \,\mathrm{yrs}}{ \sin^2 \theta_E + 0.2  \sin^2 \theta_L}\left( \frac{M_{X}/g_{5}}{10^{14}\, \mathrm{GeV}} \right)^{4}\ .
\end{align}
In order to estimate the proton lifetime, we adopt the method described in Ref.\,\cite{Nagata:2013sba}, using the hadron matrix elements of Refs.\,\cite{Aoki:2017puj,Yoo:2021gql}.
By comparing to the current limit, 
$\tau(p \rightarrow \pi^{0} + e^{+}) > 2.4\times10^{34}$\,yrs~\cite{Super-Kamiokande:2020wjk}, 
this model seemingly predicts too short proton lifetime as in the case of the conventional non-supersymmetric SU$(5)$ GUT model.
However, the proton decay width can be suppressed by the mixing angles $\theta_{E,L}$
in the fake GUT.
In the above example, 
small mixings,  $\mathrm{sin} \, \theta_{E,L} \lesssim 10^{-4}$, are consistent with the current limit on the proton lifetime. 
This suppression means that 
the SM quarks and leptons mostly have different origins.
Such small mixing angles correspond
to $m_{E,L} \ll v_{2}$, where 
the global symmetry is enhanced, as discussed in the previous subsection.

The proton lifetime in Eq.\,\eqref{eq:proton_lifetime_in_U2 }
shows that we need small mixings of both the $\overline{\mathbf{5}}_L$ and $\mathbf{10}_{\overline{E}}$ with the SM leptons to evade the constraints.
Thus, for the choices such as $H=1,~\mathrm{U}(1)_H,~\mathrm{SU}(2)_H$, for example, 
the proton lifetime is predicted to be too short unless we introduce 
a large number of $\mathrm{SU}(5)$ charged fermions~\cite{Fornal:2017xcj}.
This is the reason why $\mathrm{SU}(5)\times\mathrm{U}(2)_H$ is the minimal choice which is phenomenologically viable.%

Note that 
we estimate the above proton lifetime without considering the effects of the flavor mixings for simplicity.
In general, however, 
the lepton mass matrices in Eq.\,\eqref{eq:lepton_mass} are flavor dependent.
Hence,
an $\mathrm{SU}(5)$ gauge interaction eigenstate does not coincide with 
one generation of the SM fermions but consists of the admixture of the multiple SM generations.
Therefore, 
the predictions of the nucleon decay rates 
and the branching fractions in the fake GUT are different from those in the conventional GUT.
For example,
the decay rate of the $p \rightarrow \pi^{0} + \mu^{+}$ mode can be larger than that of the $p \rightarrow \pi^{0} + e^{+}$ mode. 
This is a contrary to the conventional GUT, where the nucleon decay modes which include different generations are suppressed by the Cabibbo--Kobayashi--Maskawa (CKM) mixing angle. 
In this way,
a variety of the nucleon decay modes will provide striking signatures of the fake GUT. 

To see the effects of the flavor dependence of the mass matrices in Eq.\,\eqref{eq:lepton_mass}, let us consider the following example of the fermion mass terms at the fake GUT scale.
\begin{align}
\label{eq: mixing parameter}
    \mathcal{M}_{L} = M_0 \begin{pmatrix}
1 & 0 & 0 & \delta_{L,11} & \delta_{L,12} & \delta_{L,13} \\
0 & 1 & 0 & \delta_{L,21} & \delta_{L,22} & \delta_{L,23} \\
0 & 0 & 1 & \delta_{L,31} & \delta_{L,32} & \delta_{L,33} 
\end{pmatrix},~~~
    \mathcal{M}_{E} = M_0 \begin{pmatrix}
1 & 0 & 0 & \delta_{E,11} & \delta_{E,12} & \delta_{E,13} \\
0 & 1 & 0 & \delta_{E,21} & \delta_{E,22} & \delta_{E,23} \\
0 & 0 & 1 & \delta_{E,31} & \delta_{E,32} & \delta_{E,33} 
\end{pmatrix}.
\end{align}
Here, we introduce small parameters $\delta_{L,E,ij}$ with $|\delta_{L,E,ij}| \ll 1$, which represents explicit breaking of the global symmetry. 
We take flavor indices of $\overline{\mathbf{5}}_i$ and $\mathbf{10}_i$ to match the generations of the quarks, and those of $L_{Hi}$ and $\overline{E}_{Hi}$ approximately corresponds to the generations of the leptons (see Eq.\,\eqref{eq:Yukawa Origins}).
In our analysis, we adopt a flavor basis that the up-type Yukawa matrix is diagonal, while the down-type Yukawa matrix has CKM-related off diagonal elements.

\begin{figure}[t!]
 	\subcaptionbox{$\delta_{L,11} = \delta_{E,11} = 10^{-4}$. \label{fig:1111}}
	{\includegraphics[width=0.46\textwidth]{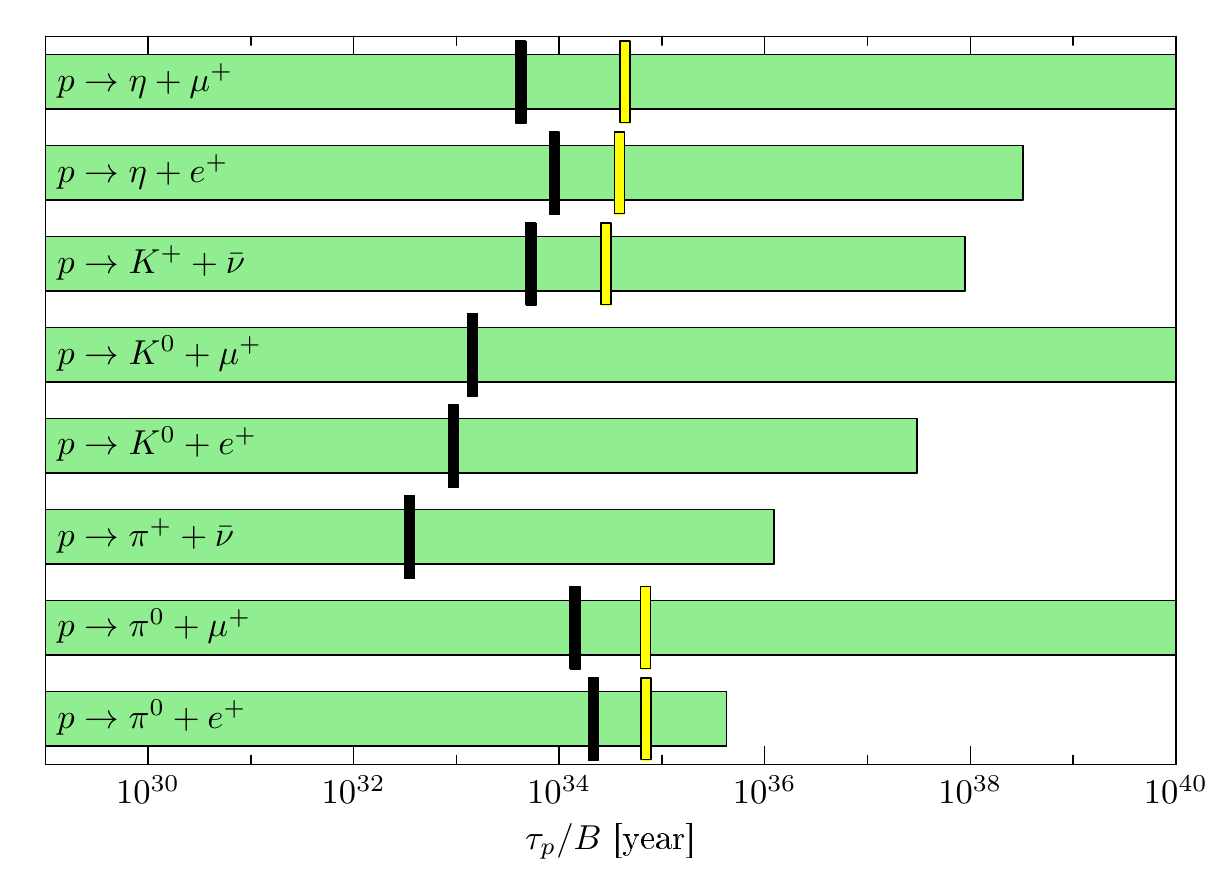}}
 	\subcaptionbox{$\delta_{L,12} = \delta_{E,12} = 10^{-4}$. \label{fig:1212}}
	{\includegraphics[width=0.46\textwidth]{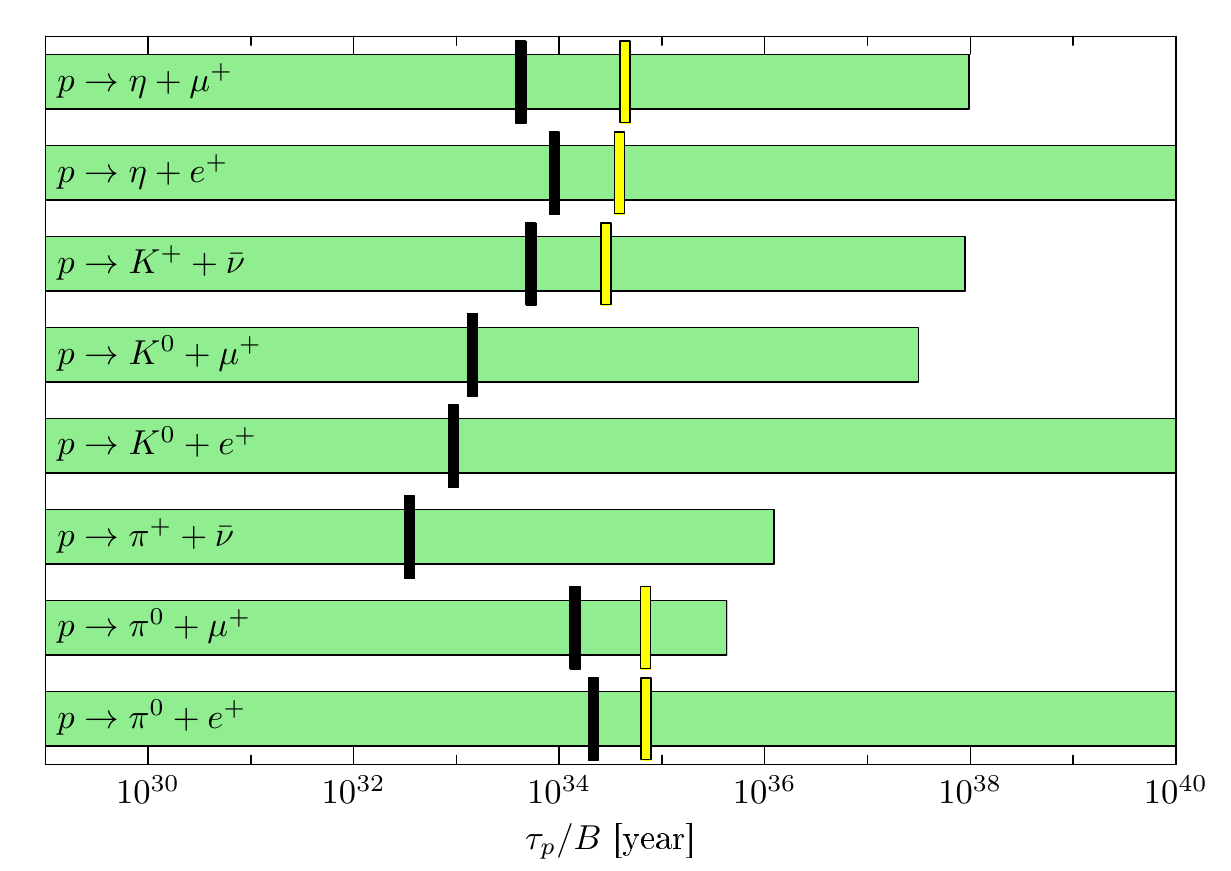}}	\\
	 \subcaptionbox{$\delta_{L,21} = \delta_{E,21} = 10^{-4}$. \label{fig:2121}}
	{\includegraphics[width=0.46\textwidth]{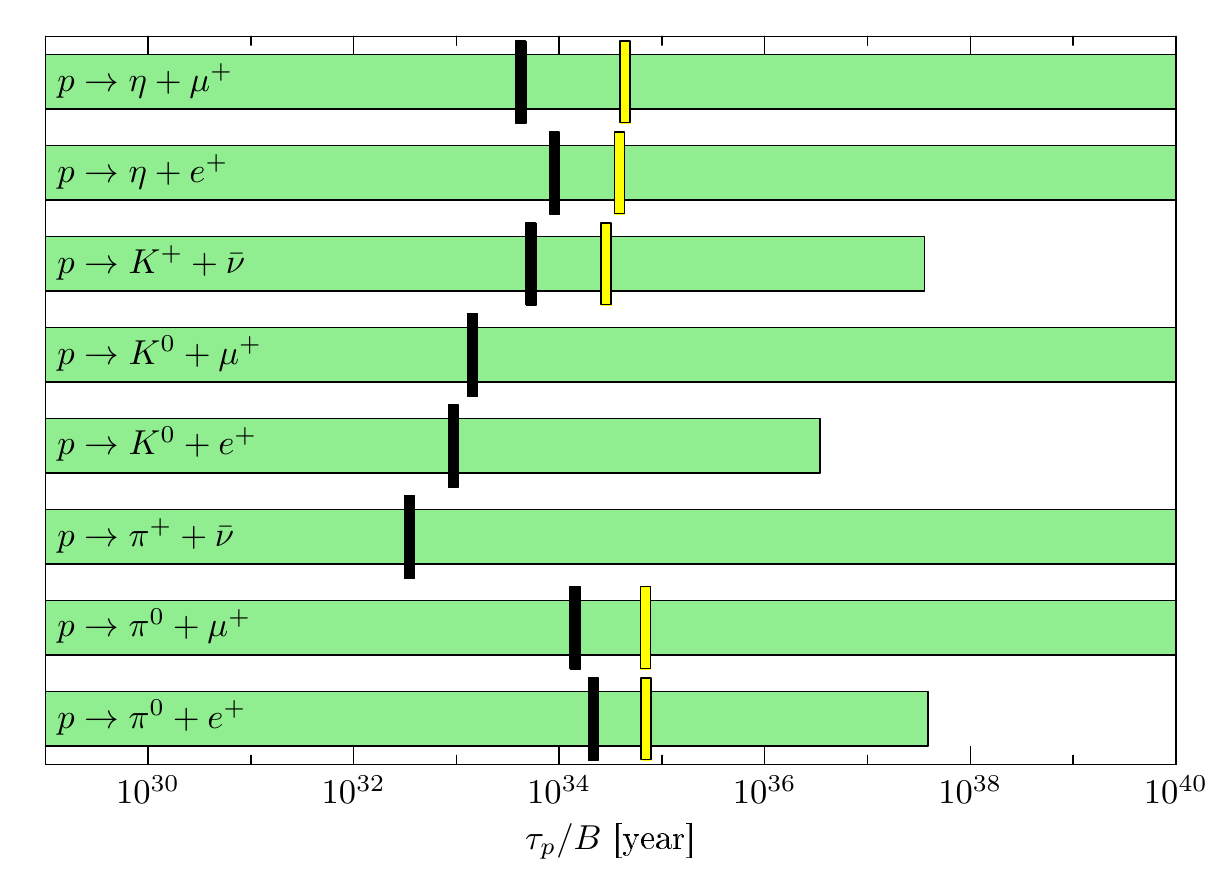}}	
 	\subcaptionbox{$\delta_{L,13} = \delta_{E,13} = 10^{-3}$. \label{fig:1313}}
	{\includegraphics[width=0.46\textwidth]{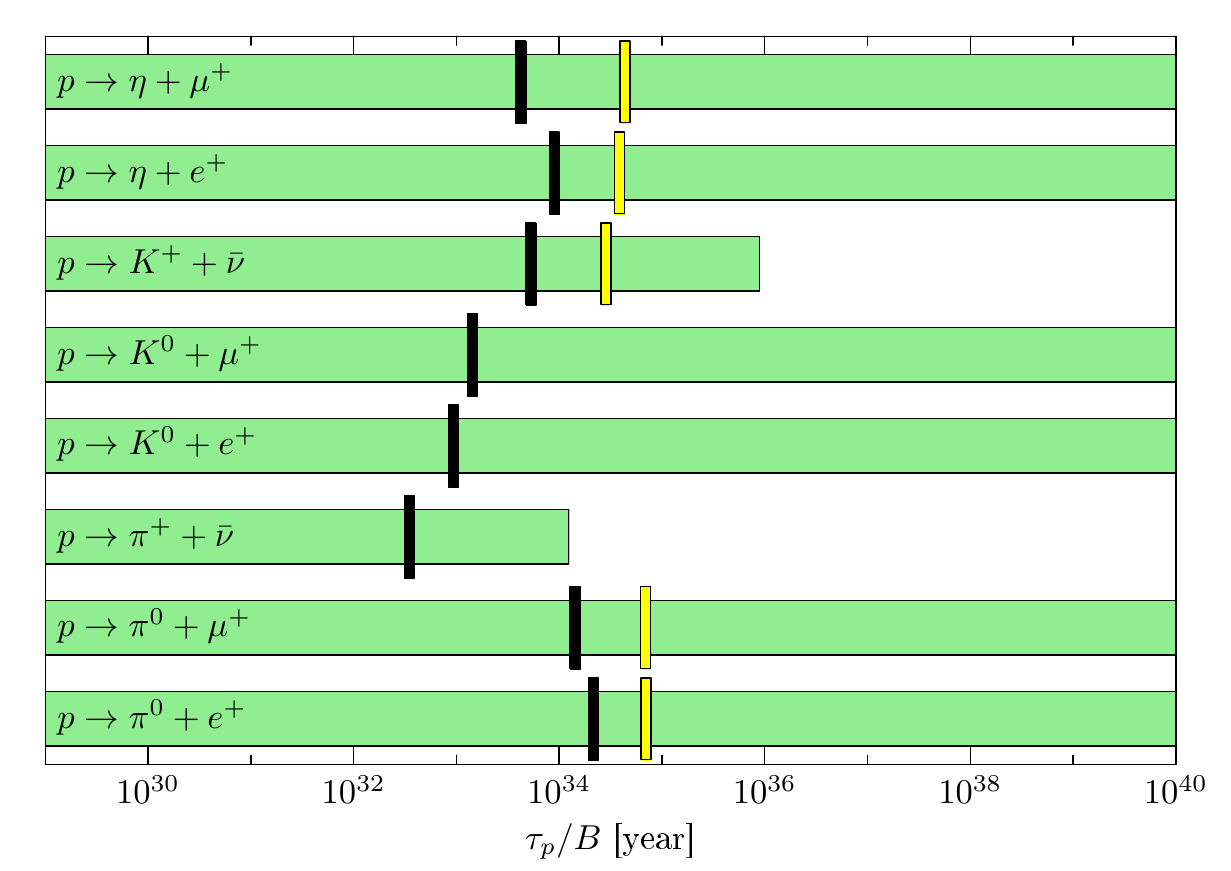}}
\caption{
The proton lifetime of each decay mode.
Here, we switch on the mixing parameters in Eq.\,\eqref{eq: mixing parameter} as indicated by the subcaptions.
We take $M_X = 10^{14}$\,GeV.
}\label{fig:decay}
\end{figure}

In Fig.\,\ref{fig:decay}, we show the proton lifetime for each mode.
In each figure, we switch on the mixing parameters 
in Eq.\,\eqref{eq: mixing parameter} as shown in the caption.
The predictions are proportional to $\delta_{L,E}^{-2}$ and $M_X^{4}$.
Black lines denote the Super-Kamiokande constraints at 90\%\,C.L. \cite{Super-Kamiokande:2005lev,Super-Kamiokande:2012zik,Super-Kamiokande:2013rwg,Super-Kamiokande:2014otb,Super-Kamiokande:2017gev,Super-Kamiokande:2020wjk} and yellow ones represent the future prospects of Hyper-Kamiokande \cite{Hyper-Kamiokande:2018ofw}.
The figures show the model predicts various leading proton decay modes depending on the mixing parameters.
For the case of the Fig.\,\ref{fig:decay}\,(b), for example,
$p\to \pi^0 + \mu^+$ mode is the leading decay mode, 
which is not expected in the conventional GUT model.

Finally, let us comment on the proton decay through the colored Higgs exchanges.
In the conventional GUT model, these contributions
is always subdominant compared with those of the 
$X$ boson exchanges due to the Yukawa suppression~\cite{Hall:2014vga},
\begin{align}
\label{eq:proton lifetime}
\tau(p\to K^+ + \nu)|_{\mathrm{colored\,Higgs \,exchange}}\sim 10^{45}\,\mathrm{yrs} \times
\theta_{L,E}^{-2}
\left(\frac{M_{H_c}} {10^{14}\,\mathrm{GeV}}\right)^4
\sin^4\theta_h
\ .
\end{align}
Here, we multiply the factor $\sin^4\theta_h$
which stems from the relative enhancement of
$y_{5,10}$, compared with the SM Yukawa couplings given in Eq.\,\eqref{eq:Yukawa Origins}.
To reproduce the masses of the SM fermions, 
we find that, the largest components of $y_{10,5,LE}$
are 
\begin{align}
 y_{10} \sim y_t/ \sin\theta_h\ , \quad 
 y_5 \sim y_b/ \sin\theta_h\ , \quad
 y_{LE} \sim y_\tau/ \cos\theta_h\ .
\end{align}
Thus, by requiring $|y_{10,5,LE}| \lesssim 1$,
\begin{align}
    \sin\theta_h \sim y_{t}/y_{10} \gtrsim 0.5 \ ,
    \quad
    \cos\theta_h \sim y_{\tau}/y_{LE} 
    \gtrsim 10^{-2}\ .
\end{align}
Therefore, there is no large enhancement from $\sin^4\theta_h$ and the proton decay rate from the colored Higgs exchange is negligible.

\section{\texorpdfstring{$\mathrm{SU}(5) \times \mathrm{SU}(3)_{H}$}{} model as UV completion of
\texorpdfstring{$\mathrm{SU}(5) \times \mathrm{U}(2)_{H}$}{} model}
\label{sec:SU3_FAKE_GUT}
In this section, we consider $\mathrm{SU}(5)\times \mathrm{SU}(3)_H$ model where $\mathrm{U}(2)_H$ is embedded into $\mathrm{SU}(3)_H$.
Note that $\mathrm{SU}(3)_H$ does not include 
$\mathrm{SU}(3)_c$ of the SM. 
As we will construct, $\mathrm{SU}(3)_H$ 
is spontaneously broken down to $\mathrm{U}(2)_H$ by the VEV of the adjoint scalar of $\mathrm{SU}(3)_H$, $A$.
The remaining $\mathrm{SU}(5)\times \mathrm{U}(2)_H$ is subsequently broken down to $G_\mathrm{SM}$ by the VEV of the scalar field in the bi-fundamental representation of $\mathrm{SU}(5)\times \mathrm{SU}(3)_H$, $\phi_3$.
That is,
\begin{align}
\label{eq:symmetry breaking path1}
\mathrm{SU}(5) \times \mathrm{SU}(3)_{H}  
\xrightarrow{\ev{A}} 
\mathrm{SU}(5) \times \mathrm{U}(2)_{H}
\xrightarrow{\ev{\phi_3}}
G_{\mathrm{SM}}.
\end{align}

In the previous section, we showed that $\mathrm{SU}(5)\times \mathrm{U}(2)_H$ model is phenomenologically viable.
However, since $H$ includes $\mathrm{U}(1)$ gauge symmetry, the model cannot explain the charge quantization unlike the conventional GUT.
The $\mathrm{U}(1)$ gauge symmetry also exhibits the Landau pole problem.
These drawbacks can be solved by the extension to $\mathrm{SU}(5)\times\mathrm{SU}(3)_H$.
We summarize the matter contents in Tab.~\ref{tab:SU3particlecontents}.

\begin{table}[t!]
\caption{The content of fermions, scalar fields and gauge bosons in the $\mathrm{SU}(5) \times \mathrm{SU}(3)_{H}$ model is shown in the group representation, $(\mathrm{SU}(5), \mathrm{SU}(3)_{H})$ and $(\mathrm{SU}(3)_c, \mathrm{SU}(2)_{L})_{\mathrm{U}(1)_{Y}}$.
Each fermion has three generations.
The superscript $R$ denotes
the real field, while the other fields are complex fields.
Some components of the fake GUT Higgs scalars $\phi_3$ and $A$ become the longitudinal modes
of the fake GUT gauge boson when $\mathrm{SU}(5)\times \mathrm{SU}(3)_H$ 
is broken to $G_\mathrm{SM}$ (see text).
}
  \label{tab:SU3particlecontents}
 \begin{center}
  \begin{tabular}{|c||c|c|}  \hline
     & $(\mathrm{SU}(5),\mathrm{SU}(3)_{H})$ & $(\mathrm{SU}(3)_{c},\mathrm{SU}(2)_{L})_{\mathrm{U}(1)_{Y}}$ \\ \hline \hline
    fermions &  &  \\ \hline
    $\overline{\mathbf{5}}$ & $(\overline{\mathbf{5}},\bf{1})$ & $(\overline{\mathbf{3}},\bf{1})_{\mathrm{+1/3}} \oplus (\mathbf{1},\mathbf{2})_{\mathrm{-1/2}}$ \\ 
    $\mathbf{10}$ & $(\mathbf{10},\bf{1})$ & $(\mathbf{3},\mathbf{2})_{\mathrm{+1/6}} \oplus (\overline{\mathbf{3}},\mathbf{1})_{\mathrm{-2/3}} \oplus (\mathbf{1},\mathbf{1})_{\mathrm{+1}}$ \\
    $L_{T}$ & $(\mathbf{1},\mathbf{3})$ & 
    $(\mathbf{1},\mathbf{2})_{\mathrm{-1/2}} \oplus (\mathbf{1},\mathbf{1})_{\mathrm{+1}} $\\
    $\overline{L}_{T}$ & $(\mathbf{1},\mathbf{\overline{3}})$ & 
    $(\mathbf{1},\mathbf{2})_{\mathrm{+1/2}} \oplus (\mathbf{1},\mathbf{1})_{\mathrm{-1}}$\\ \hline \hline
    scalars &  &  \\ \hline
    $A$ & $(\mathbf{1}, \mathbf{8})$ & $(\mathbf{1},\mathbf{3})_{\mathrm{0}}^{R} \oplus (\mathbf{1},\mathbf{2})_{\mathrm{+3/2}} \oplus  (\mathbf{1},\mathbf{1})_{\mathrm{0}}^{R} $ \\
    $\phi_{3}$ & 
    $(\mathbf{5}, \mathbf{3})$
    & $(\mathbf{3},\mathbf{2})_{\mathrm{-5/6}} \oplus 
    (\mathbf{3},\mathbf{1})_{\mathrm{+2/3}} \oplus
    (\mathbf{1},\mathbf{2})_{\mathrm{+3/2}} \oplus
    (\mathbf{1},\mathbf{3})_{\mathrm{0}} \oplus
    (\mathbf{1},\mathbf{1})_{\mathrm{0}} $
    \\
    $H_{5}$ & $(\mathbf{5}, \mathbf{1})$ 
    & $(\mathbf{3},\mathbf{1})_{\mathrm{-1/3}} \oplus (\mathbf{1},\mathbf{2})_{\mathrm{+1/2}}$ \\
    $H_{3}$ & 
    $(\mathbf{1}, \mathbf{3})$ & 
    $(\mathbf{1},\mathbf{2})_{\mathrm{-1/2}} \oplus (\mathbf{1},\mathbf{1})_{\mathrm{+1}} $ \\ \hline \hline
    vectors &  &  \\ \hline
    $V_{5}$ & $(\mathbf{24}, \mathbf{1})$ & $(\mathbf{8},\mathbf{1})_{\mathrm{0}}^{R} \oplus (\mathbf{3},\mathbf{2})_{\mathrm{-5/6}} \oplus  (\mathbf{1},\mathbf{3})_{\mathrm{0}}^{R} \oplus (\mathbf{1},\mathbf{1})_{\mathrm{0}}^{R}$ \\
    $V_{3H}$ & $(\mathbf{1}, \mathbf{8})$ & $(\mathbf{1},\mathbf{3})_{\mathrm{0}}^{R} \oplus (\mathbf{1},\mathbf{2})_{\mathrm{+3/2}} \oplus  (\mathbf{1},\mathbf{1})_{\mathrm{0}}^{R} $ \\ \hline
  \end{tabular}
 \end{center} 
\end{table}

\subsection{Spontaneous symmetry breaking of \texorpdfstring{$\mathrm{SU}(5) \times \mathrm{SU}(3)_{H}$}{}}
\label{sec:Model of NON ABELIAN FAKE GUT}
In this model,
we introduce a real scalar field, $A$, in  $(\mathbf{1},\mathbf{8})$
representation, and a complex scalar field, $\phi_3$, in
$(\mathbf{5},\mathbf{3})$  representation of $(\mathrm{SU}(5), \mathrm{SU}(3)_{H})$, respectively.
The scalar potential is 
\begin{align}
\label{eq:phi,A pote}
V(A, \, \phi_3)
&= -\mu^{2}_{A} \mathrm{Tr} \left[ A^{2} \right]
+ \lambda_{1 A} \left( \mathrm{Tr} \left[ A^{2} \right] \right)^{2}
+ \lambda_{2 A} \mathrm{Tr} \left[ A^{4} \right]
+ \mu_{3 A} \mathrm{Tr} \left[ A^{3} \right] \nonumber \\
&- \mu^{2}_{\phi} \mathrm{Tr} \left[ \phi_3^{\dagger} \phi_3 \right]
+ \lambda_{1 \phi} \left( \mathrm{Tr} \left[ \phi_3^{\dagger} \phi_3 \right] \right)^{2}
+ \lambda_{2 \phi} \mathrm{Tr} \left[ \phi_3^{\dagger} \phi_3 \phi_3^{\dagger} \phi_3 \right] \nonumber \\
&+ \lambda_{1 \phi A} \mathrm{Tr} \left[ \phi_3^{\dagger} \phi_3 \right] \mathrm{Tr} \left[ A^{2} \right]
+ \lambda_{2 \phi A} \mathrm{Tr} \left[ \phi_3^{\dagger} A^{2} \phi_3  \right]
- \mu_{3 \phi A} \mathrm{Tr} \left[ \phi_3^{\dagger} A \phi_3  \right].
\end{align}
Here, $\mu$'s are parameters with mass dimension one, and $\lambda$'s are dimensionless parameters.
We take all the parameters are positive for simplicity.
The vacuum in $\mathrm{SU}(5)\times \mathrm{U}(2)_H$ model in Eq.\,\eqref{eq: Phi vev}
is reproduced when 
$A$ and $\phi_3$ take the VEVs
in the following form;
\begin{align}
\label{eq:SU(3) breaking}
\ev{A} 
= \left( \begin{array}{ccc}
      v_{A} & 0 & 0 \\
      0 & v_{A} & 0 \\
      0 & 0 & -2v_{A}
    \end{array} \right), \,\,\,
\ev{\phi_3} 
= \left( \begin{array}{ccccc}
      0 & 0 & 0 & v_{3} & 0 \\
      0 & 0 & 0 & 0 & v_{3} \\
      0 & 0 & 0 & 0 & 0
    \end{array} \right).
\end{align}
To achieve this form, 
$v_A\neq 0$ is crucial, 
since otherwise we find that either 
only a single row of $\ev{\phi_3}$ or all the three rows of $\ev{\phi_3}$ are non-vanishing.%
\footnote{The symmetry is broken as $\mathrm{SU}(5)\times \mathrm{SU}(3)_H \to
\mathrm{SU}(4)\times \mathrm{SU}(2)\times \mathrm{U}(1)$
for for the former case,
and $\mathrm{SU}(5)\times \mathrm{SU}(3)_H \to
\mathrm{SU}(3)\times \mathrm{SU}(2)$ for the latter case.}
Once $v_A \neq 0$, 
the last two terms in Eq.\eqref{eq:phi,A pote}
can make the mass squared of the third row of $\phi_3$ positive while those of the first and the second rows are kept negative.
For $v_A \gtrsim v_3$, 
we may regard that 
the symmetry breaking takes place 
in the two steps as follows;
\begin{align}
\label{eq:symmetry breaking path2}
\mathrm{SU}(5) \times \mathrm{SU}(3)_{H}  
\xrightarrow{\ev{A}} 
\mathrm{SU}(5) \times \mathrm{U}(2)_{H}
\xrightarrow{\ev{\phi_3}}
G_{\mathrm{SM}},
\end{align}
where the second step corresponds to the symmetry breaking in the $\mathrm{SU}(5)\times \mathrm{U}(2)_H$ model.
Note that the model does not require the hierarchy between $v_A$ and $v_3$, 
and it is possible to consider situation where 
$\mathrm{SU}(5)\times \mathrm{SU}(3)_H$ 
breaks down directly to $G_\mathrm{SM}$ 
for $v_A < v_3$.

In Tab.\,\ref{tab:SU3vectormass}, we show the 
gauge boson masses.
The $X$ boson absorbs $(\mathbf{3},\mathbf{2})_{\mathrm{-5/6}}$ scalar
in Tab.\,\ref{tab:SU3particlecontents}
as a longitudinal mode.
The $\Omega_{2}$ boson absorbs a linear combination of two $(\mathbf{1}, \mathbf{2})_{+3/2}$ scalar fields.
The $\Omega_{3}$  ($\Omega_{1}$) 
absorb 
a linear combination of  $(\mathbf{1},\mathbf{3})_0$ 
($(\mathbf{1},\mathbf{1})_0$) 
in the bi-fundamental fake GUT Higgs $\phi_3$.
Accordingly, the physical fake GUT Higgs bosons
appearing from $\phi_3$ and $A$
consist of a $(\mathbf{1}, \mathbf{2})_{+3/2}$ scalar field, 
two $(\mathbf{1},\mathbf{3})^R_0$ scalars,
two $(\mathbf{1},\mathbf{1})^R_0$ scalars,
and a $(\mathbf{3},\mathbf{1})_{+2/3}$ scalar.

\begin{table}[tb]
\caption{The mass spectrum of the gauge bosons.
The $X_{\mu}$ is the $\mathrm{SU}(5)$ gauge fields.
$\Omega_{2\mu}$, $\Omega_{3\mu}$ and $\Omega_{1\mu}$ are the $\mathrm{SU}(3)_{H}$ gauge fields.
}
 \begin{center}
  \begin{tabular}{|c||c|c|}  \hline
    vectors & $(\mathrm{SU}(3)_{c},\mathrm{SU}(2)_{L})_{\mathrm{U}(1)_{Y}}$ & mass \\ \hline \hline
    $X_{\mu}$ & $(\mathbf{3},\mathbf{2})_{\mathrm{-5/6}}$ & $g_{5}v_{3}/\sqrt{2}$ \\
    $\Omega_{2\mu}$ & $(\mathbf{1}, \mathbf{2})_{+3/2}$ & $3g_{3H} \sqrt{v_{A}^{2}+v_{3}^{2}/18}$ \\
    $\Omega_{3\mu}$ & $(\mathbf{1}, \mathbf{3})_{0}$ & $\sqrt{g_{5}^{2} + g_{2H}^{2}} v_{3}$ \\
    $\Omega_{1\mu}$ & $(\mathbf{1}, \mathbf{1})_{0}$ & $\sqrt{3g_{5}^{2}/5 + g_{1H}^{2}} v_{3}$ \\ \hline
  \end{tabular}
  \label{tab:SU3vectormass}
 \end{center} 
\end{table}

In the previous $\mathrm{SU}(5)\times \mathrm{U}(2)_{H}$ model,
the matching conditions between the SM and $\mathrm{SU}(5)\times \mathrm{U}(2)_{H}$ gauge couplings
are given in Eq.\,\eqref{eq:matching}.
For $v_A \gtrsim v_3$
the matching conditions 
of the gauge coupling constants 
at the symmetry breaking in the second step
in Eq.\,\eqref{eq:symmetry breaking path2}
are also given by Eq.\,\eqref{eq:matching} 
with $M_X$ replaced by $M_X = g_5 v_3/\sqrt{2}$.

The matching conditions between $\mathrm{SU}(5)\times \mathrm{U}(2)_{H}$ and $\mathrm{SU}(5)\times\mathrm{SU}(3)_{H}$ models are, on the other hand, given by
\begin{align}
\label{eq:SU3,U1matching}
\begin{split}
\frac{1}{3} \alpha_{1H}^{-1} (M_{\Omega})
&= \alpha_{3H}^{-1}(M_{\Omega}), \\
\alpha_{2H}^{-1}(M_{\Omega}) &= \alpha_{3H}^{-1}(M_{\Omega})\ .
\end{split}
\end{align}
Here, $M_{\Omega} = 3g_{3H}\sqrt{v_A^2 + v_3^{2}/18}$ is the energy scale at which these two models are matched.
The factor $1/3$ in Eq.\,\eqref{eq:SU3,U1matching} appears since the $\mathrm{U}(1)_H$ charge
is embedded in the fundamental representation of $\mathrm{SU}(3)_H$ with the normalization, $\tr[t^a t^b]=\delta^{ab}/2$, with $t^a$ being the generator of $\mathrm{SU}(3)_H$.\footnote{The generator of $\mathrm{U}(1)_H$ embedded in $\mathrm{SU}(3)_H$
corresponds to $-t_8$ with $t_8$
being the eighth Gell-Mann matrix divided by $2$.
}

\begin{figure}[t!]
	\centering
 	\subcaptionbox{$M_X = 10^{7}$\,GeV \label{fig:1e7GeV}}
	{\includegraphics[width=0.46\textwidth]{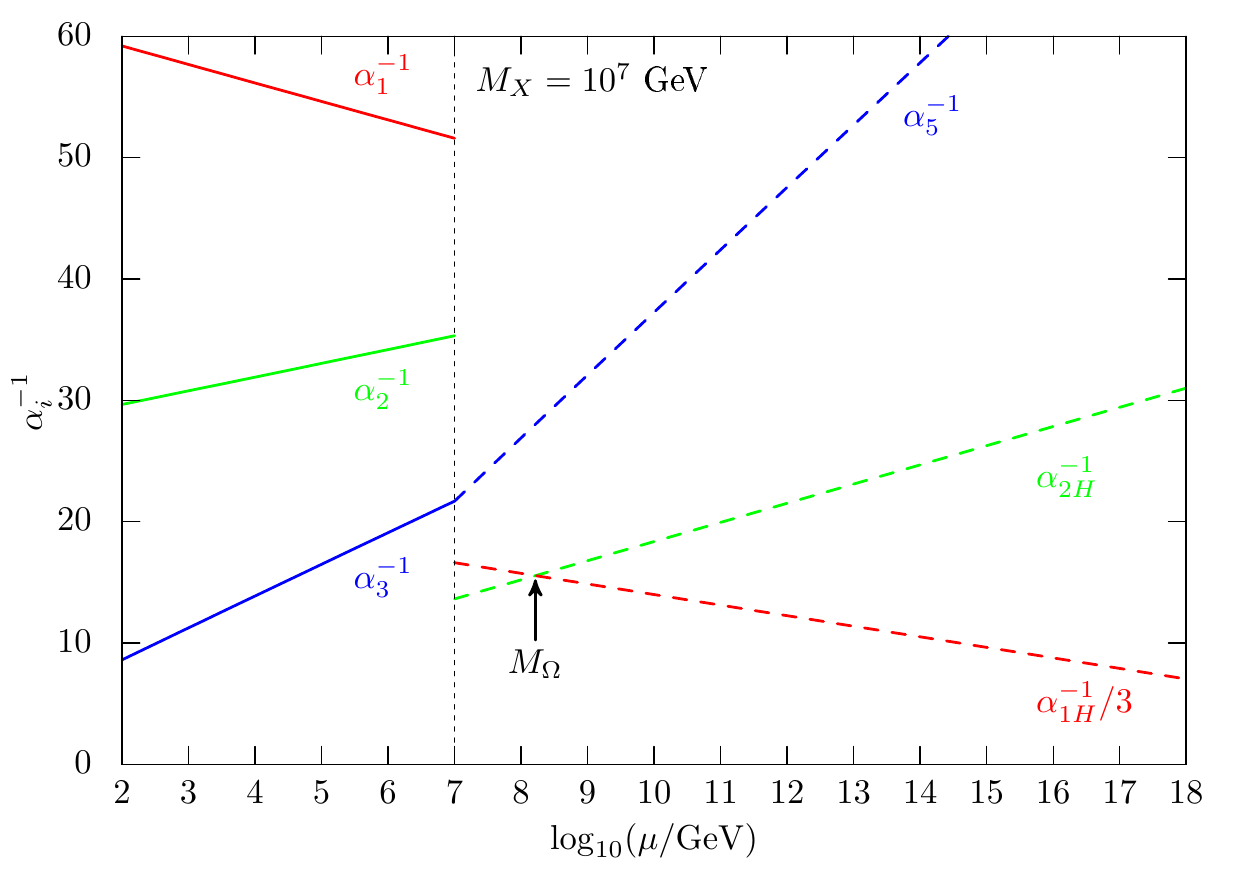}}
 	\subcaptionbox{$M_X = 10^{13}$\,GeV\label{fig:1e13GeV} }
	{\includegraphics[width=0.46\textwidth]{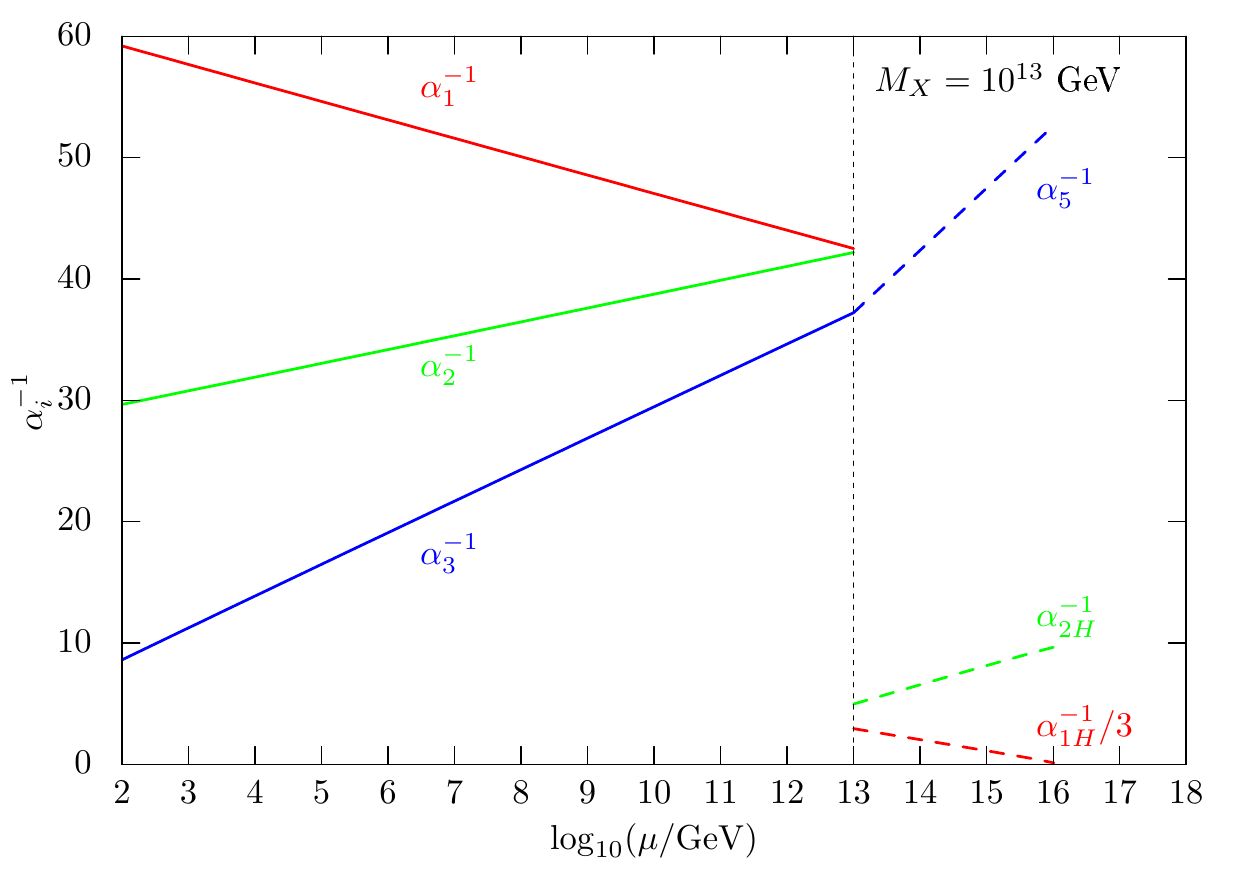}}
\caption{
The left (right) figure shows the running of gauge couplings when $M_{X} =  10^{7} \, \mathrm{GeV}$ ($M_{X} =10^{13} \, \mathrm{GeV}$). 
The red, green, and blue solid lines denote the runnings of $\alpha^{-1}_1,~\alpha_2^{-1}$, and $\alpha_3^{-1}$, respectively.
The red, green, and blue dashed lines are, respectively, the runnings of $\alpha^{-1}_{1H}/3,~\alpha^{-1}_{2H}$, and $\alpha_5^{-1}$.
Here we assume that only the particles contained in the $\mathrm{SU}(5)\times \mathrm{U}(2)_H$ model contribute to the running of gauge couplings.
The figure shows that the matching between $\mathrm{U}(2)_H$ 
and $\mathrm{SU}(3)_H$ is possible for the left panel while it is not possible in the right panel.
}\label{fig:2loop}
\end{figure}

For a given $M_X$, the gauge couplings $\alpha_{1H}$ and $\alpha_{2H}$ are determined by Eq.\,\eqref{eq:matching}. 
Above $M_X$, 
$\alpha_{5}$, 
$\alpha_{1H}$, and $\alpha_{2H}$
evolves following 
the renormalization group (RG) equations in 
the $\mathrm{SU}(5)\times \mathrm{U}(2)_H$ model for $v_A\gtrsim v_3$.
The RG equations of the present model are given in the Appendix~\ref{sec:RGE}.
We show examples of the RG running in Fig.\,\ref{fig:2loop}.
By using these running gauge couplings, we can find the matching scale $M_\Omega^{\mathrm{match}}$ which satisfies 
the conditions in Eq.\,\eqref{eq:SU3,U1matching}.

For $v_A \lesssim v_3$, all the gauge boson masses are dominated by $v_3$ contribution, and hence, $M_X \simeq M_\Omega$.
Accordingly, the matching conditions in Eqs.\,\eqref{eq:matching} and \eqref{eq:SU3,U1matching} are combined where the matching conditions are given at the same scale, $M_X \simeq M_\Omega$.

\begin{figure}[t!]
	\centering
	\includegraphics[width=0.6\textwidth]{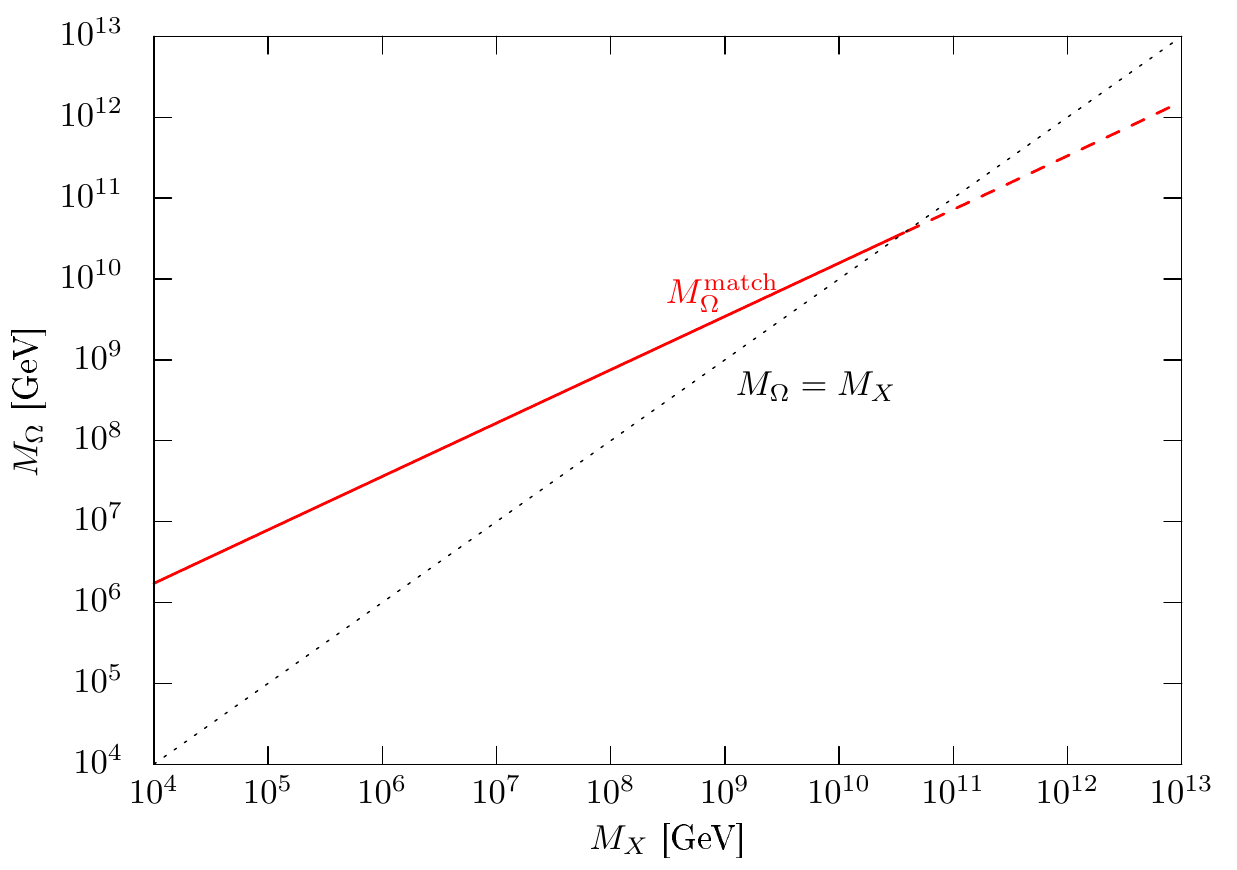}
\caption{
The matching scale $M_\Omega^{\rm match}$ as a function of $M_X$ (red solid line).
}\label{fig:match}
\end{figure}

Since $M_\Omega \gtrsim M_X$ in the present model,
we find that there is an upper limit on $M_X$ which can be seen from Fig.\,\ref{fig:match};
\begin{align}
\label{eq:M2 condition}
M_{X} 
\le 4 \times 10^{10}\,\mathrm{GeV}\ .
\end{align}
Due to this severer constraints on $M_{X}$ than that in $\mathrm{SU}(5)\times \mathrm{U}(2)_H$ model, $M_X \lesssim 10^{14}\,$GeV, (see Fig.\,\ref{fig:SM_running}), 
this model requires smaller mixing angles of the lepton components of $\overline{\mathbf{5}}\oplus \mathbf{10}$ to evade the constraints from the proton lifetime;
\begin{align}
\label{eq:limit on angle}
    \theta_{L,E} \lesssim 10^{-12} \left(\frac{M_X/g_5}{10^{10}\,\mathrm{GeV}}\right)^2\ ,
\end{align}
(see Eq.\,\eqref{eq:proton_lifetime_in_U2 }).
To achieve the highly suppressed lepton mixing angle, we need a high quality lepton symmetry.
In the App.\,\ref{sec:On global lepton and baryon symmetry},
we give an example of a model in which such a high quality lepton symmetry originates from a discrete gauge symmetry.

\subsection{SM fermions, SM Higgs and Yukawa interactions}
\label{sec: scale in SU(3) model}
In the $\mathrm{SU}(5)\times \mathrm{U}(2)_H$ model,
the vector-like fermions are charged under $H=\mathrm{U}(2)_H$ are $(L_{H}, \overline{L}_{H})$ and $(E_{H}, \overline{E}_{H})$.
In the $\mathrm{SU}(5)\times \mathrm{SU}(3)_H$ model,
they are straightforwardly embedded into the $\mathrm{SU}(3)_{H}$ fundamental
representations, 
\begin{align}
(\,L_{T} 
= ({L}_{H} \, \overline{E}_{H}) : (\mathbf{1},\mathbf{3}), \,\,\, 
\overline{L}_{T} 
= (\overline{L}_{H} \, {E}_{H}) :(\mathbf{1},\mathbf{\overline{3}})\,)\ \, \times 3.
\end{align}
As in the case of $\mathrm{SU}(5)\times \mathrm{U}(2)_H$ model,
the leptonic components, $\overline{\mathbf{5}}_L$ and $\mathbf{10}_E$
become the mass partners of $L_{T}$ and $\overline{L}_{T}$ through the coupling 
to $\phi_{3}$,
\begin{align}
\label{eq:su(3) mass}
\mathcal{L} 
= m_{T} \overline{L}_{T} L_{T}
+ \lambda_{5} \overline{L}_{T} \phi_3 \overline{\mathbf{5}}
+ \frac{\lambda_{10}}{\Lambda_{\mathrm{cut}}} \overline{L}_{T} \phi_3^{\dagger} \phi_3^{\dagger} \mathbf{10}
+\lambda_{TAT} \overline{L}_{T} A L_{T} + h.c..
\end{align}
Here, $m_{T}$ is a mass parameter of the vector-like fermions and $\lambda$'s are dimensionless couplings. 
The flavor indices are omitted for simplicity.
The $\mathrm{SU}(3)_H$ indices of $\overline{L}_T\phi_3^\dagger \phi_3^\dagger$
are contracted by the totally anti-symmetric invariant tensor of $\mathrm{SU}(3)_H$,
while its $\mathrm{SU}(5)$ indices are 
contracted with those of $\mathbf{10}$ which is the anti-symmetric representation. 
The cutoff scale $\Lambda_{\mathrm{cut}}$
is a scale larger than the fake GUT scale $v_3$ and $v_A$ corresponding to some heavier particles.
An example of such particles is Dirac fermion with $(\mathbf{5},\overline{\mathbf{3}})$.
We may also consider a complex scalar
$\phi_3'$ in $(\overline{\mathbf{10}},{\mathbf{3}})$ 
which has a Yukawa coupling 
$\overline{L}_T\phi_3' \mathbf{10}$ and 
a trilinear coupling $\phi_3 \phi_3' \phi_3$ (see the App.~\ref{sec:On global lepton and baryon symmetry}
for a concrete model.).

As in the previous section,
the SM quarks originate from 
$\overline{\mathbf{5}}\oplus\mathbf{10}$.
The lepton components in $\overline{\mathbf{5}}$ 
and $\mathbf{10}$ and $L_{T}$'s
obtain $3\times 6$ mass matrices from the interactions in Eq.\,\eqref{eq:su(3) mass}, and 
the three leptons remain massless due to the rank conditions.
Since 
the $\overline{\mathbf{5}}\oplus\mathbf{10}$
contributions to the SM leptons are 
highly suppressed (see Eq.\,\eqref{eq:limit on angle}), we require
$m_T \ll v_{3,A}$ and $\lambda_{TAT}\ll 1$.
As in the case of the $\mathrm{SU}(5)\times \mathrm{U}(2)_H$ model, 
the global $\mathrm{U}(1)_{LT}$ symmetry is
enhanced in addition to  
the global $\mathrm{U}(1)_5$ symmetry
in the limit of vanishing $m_T$ and $\lambda_{TAT}$.
In the App.~\ref{sec:On global lepton and baryon symmetry}, we discuss a model where the lepton symmetry originates from a discrete gauge symmetry.

Next, we discuss the origin of the SM Higgs and Yukawa interactions in the $\mathrm{SU}(5) \times \mathrm{SU}(3)_{H}$ model.
As in the $\mathrm{SU}(5) \times \mathrm{U}(2)_{H}$ model, the SM Yukawa interactions come from various contributions, and we consider a case that only one $\mathrm{SU}(2)_L$ Higgs doublet remains in the low energy.
We introduce $H_{5}$ and $H_{3}$ scalar fields, which are  $(\mathbf{5}, \mathbf{1})$ and $(\mathbf{1}, \mathbf{3})$ representations.
The $H_3$ and $H_5$ are decomposed as,
\begin{align}
    H_3 =   \left(
\begin{array}{c}
     h_3^{\mathrm{SM}\dagger}  \\
     h_3^{\mathrm{singlet}} 
\end{array}
    \right) \ , \quad H_5 = 
    \left(
\begin{array}{c}
     h_5^{\mathrm{color}}  \\
     h_5^{\mathrm{SM}} 
\end{array}
    \right)\ ,
\end{align}
where a linear combination of $h_{3}^{\mathrm{SM}}$ and $h_{5}^{\mathrm{SM}}$ 
becomes the SM Higgs doublet.
The mixing term between $h_3^\mathrm{SM}$ and 
$h_5^\mathrm{SM}$ comes from
\begin{align}
\mathcal{L}_{53 \mathrm{mix}} 
= \mu_{\phi 5 3 } \, H^{\dagger}_{3} \phi_{3} H_{5}^{\dagger}  + h.c.,
\end{align}
where $\mu_{\phi 53}$ is the mass parameter of the order of the fake GUT scale.
The SM Higgs doublet in the electroweak scale requires fine-tuning.
The singlet and the colored Higgs as well as a heavier combination of $h_{3}^\mathrm{SM}$ and $h_5^{\mathrm{SM}}$ obtain masses of the order of the fake GUT scale.

The Yukawa interactions of the quarks are given by those of $\mathrm{SU}(5)$ multiplets 
with $H_5$ as in Eq.\,\eqref{eq:quark Yukawa}.
Since the SM leptons dominantly come from 
$L_T$'s, we expect that the SM lepton Yukawa couplings are dominated by the couplings to $H_3$, 
\begin{align}
\label{eq:SU3Yukawa}
\mathcal{L}_{YL} 
= - (y_{LT})_{ij} \epsilon_{\alpha \beta \gamma} H_{3}^{\alpha} {L}_{Ti}^{\beta} {L}_{Tj}^{\gamma} + h.c.,
\end{align}
which is the naive extension of Eq.\,\eqref{eq:lepton Yukawa}.
Here, the subscript $\alpha$ is the index of the $\mathrm{SU}(3)_{H}$ and $i$ and $j$ run the number of generations.
Since $\epsilon_{\alpha \beta \gamma}$ is totally antisymmetric, 
we find
$(y_{LT})_{ij} =-(y_{LT})_{ji} $. 
The 
3-by-3 antisymmetric Yukawa coupling results in
the massless electron and the $\mu$ and $\tau$ lepton in the same mass.
Thus, the Yukawa coupling given by Eq.\,\eqref{eq:SU3Yukawa} does not reproduce the lepton masses in the SM.

As in Eq.\,\eqref{eq:Yukawa Origins},
the lepton Yukawa couplings receive the contribution from $y_5$, which are suppressed by the lepton mixing angles $\theta_{L}\times \theta_E$.
The lepton mixing angles are, however, required 
to be highly suppressed, $\theta_{L,E}\lesssim 10^{-12}$, to evade the constraints from the 
proton lifetime (see Eq.\,\eqref{eq:limit on angle}). 
Thus, those contributions are too small to 
reproduce the lepton masses in the SM,
and hence, we need other origins of the lepton Yukawa interactions.

As an example to generate appropriate lepton Yukawa coupling, we consider the following higher dimensional operator,
\begin{align}
\label{eq:su3_Yukawa}
\mathcal{L}_{YL} 
= - \frac{(Y_{LT})_{ij}}{\Lambda_{Y}} {L}_{Ti}^{\alpha} A_{\alpha}^{\beta} {L}_{Tj}^{\gamma} H_{3}^{\delta} \epsilon_{\beta \gamma \delta}
+ h.c.,
\end{align}
where $\Lambda_{Y}$ is a cutoff scale.
The scale $\Lambda_{Y}$ can be given by the mass of heavy particles such as Dirac fermions in 
$(\mathbf{1}, \mathbf{3})$ representation (see App.~\ref{sec:On global lepton and baryon symmetry}).
Once $A$ obtains the VEV, this operator generates 
the lepton Yukawa couplings.
Unlike the Yukawa couplings in Eq.\,\eqref{eq:SU3Yukawa}, the 3-by-3 coefficient matrix is not anti-symmetric, 
and hence, this operator can reproduce the SM lepton spectrum, when $\Lambda_Y$ is smaller than $\order{10^2}\times v_A$.

\section{Conclusions and discussions}
\label{sec:conclusions}

The fake GUT is a framework which has been proposed to explain the perfect fit of the SM matter fields into the $\mathrm{SU}(5)$ multiplets \cite{Ibe:2019ifm}.
Unlike the conventional GUT \cite{Georgi:1974sy}, 
the quarks and leptons are not necessarily embedded in common 
GUT multiplets but embedded in different multiplets although 
they appear to form complete GUT multiplets at the low energy.
In this paper, we studied details of the model based on $\mathrm{SU}(5) \times \mathrm{U}(2)_{H}$ gauge symmetry and its extension 
with $\mathrm{SU}(5) \times \mathrm{SU}(3)_{H}$ gauge symmetry.
We discussed the nature of the fake GUT symmetry breaking.
We also studied the origins of the SM quarks/leptons, the Higgs fields, and the SM Yukawa interactions.

In the present models, the global lepton and baryon symmetries play a crucial role to avoid too rapid nucleon decays. 
However, these global symmetries are less likely exact ones, due to theoretical and phenomenological reasons.
With the violation of the global symmetries, the nucleons are no longer stable.
The decay rates and decay modes strongly depend on its size and  flavor structure.
Observations of multiple nucleon decay modes are the smoking guns of the present models.

We extended the $\mathrm{SU}(5) \times \mathrm{U}(2)_{H}$ model to the $\mathrm{SU}(5) \times \mathrm{SU}(3)_{H}$ model to explain the $\mathrm{U}(1)_{H}$ charge quantization and avoid Landau pole problem.
As for $\mathrm{SU}(3)_H$ model,
we discussed the scalar potential 
with which the $\mathrm{SU}(2)_L$ appears in the diagonal subgroup of $\mathrm{SU}(5)\times \mathrm{SU}(3)_H$.
As a result, we find that the model with the scalar fields in the bi-fundamental representation of $\mathrm{SU}(5)\times \mathrm{SU}(3)_H$
and in the adjoint representation of $\mathrm{SU}(3)_H$
provides the successful symmetry breaking.
We also studied the upper limit on the fake GUT scale, $M_X \lesssim 10^{10}$\,GeV.
Due to the low fake GUT scale, 
the model requires rather strict lepton symmetry to suppress the proton decay.

Let us briefly discuss alternative extension of $\mathrm{SU}(5)\times \mathrm{U}(2)_H$ model,  
instead of $\mathrm{SU}(5)\times \mathrm{SU}(3)_H$.
For example, one may consider models based on the 
gauge groups $\mathrm{SU}(6)\times \mathrm{SU}(2)_H
\to \mathrm{SU}(5)\times \mathrm{U}(2)_H$ 
or $\mathrm{SU}(7)\to \mathrm{SU}(5)\times \mathrm{U}(2)_H$.
Those groups have chiral representations $\overline{\mathbf{6}}\oplus \overline{\mathbf{6}}\oplus \mathbf{15}$
for  $\mathrm{SU}(6)$
and $\overline{\mathbf{7}}\oplus\overline{\mathbf{7}}\oplus \overline{\mathbf{7}}\oplus \mathbf{21}$
for  $\mathrm{SU}(7)$, respectively.
However, the resultant $\mathrm{SU}(5)\times \mathrm{U}(2)_H$ models do not satisfy the conditions of the fake GUT model, and hence, they do not provide viable UV completions of the $\mathrm{SU}(5)\times \mathrm{U}(2)_H$ model.

Finally let us comment on the topological objects generated in the early Universe in these models.
Unlike the conventional GUT based on simple groups, the $\mathrm{SU}(5) \times \mathrm{U}(2)_{H}$ model has no dangerous topological objects.
Therefore high-scale inflation and high reheating temperature are allowed in the model.
In the case of $\mathrm{SU}(5) \times \mathrm{SU}(3)_{H}$, on the other hand, it is possible to generate monopoles 
if the fake GUT breaking takes place after the inflation.
In order to avoid this monopole problem, the reheating temperature should be much less than the fake GUT scale.
Detailed constraints 
and the possible detection of the monopole will be explored elsewhere.
Besides, let us also comment on the monopole catalysis of proton decay. In the conventional GUT model, the monopoles induce the baryon-number non-conserving reactions~\cite{Rubakov:1982fp,Callan:1982ah}.
In the fake GUT model, on the other hand, baryon-number violating processes, $e.g.$ proton decay, are also controlled by the lepton symmetry unlike the conventional GUT. This may provide different features of the monopole in the fake GUT model compared to that in the conventional GUT.

\section*{Acknowledgements}
This work is supported by Grant-in-Aid for Scientific Research from the Ministry of Education, Culture, Sports, Science, and Technology (MEXT), Japan, 18H05542, 21H04471, 22K03615 (M.I.), 18K13535, 20H01895, 20H05860 and 21H00067 (S.S.) and by World Premier International Research Center Initiative (WPI), MEXT, Japan. 
T.T.Y. is supported in part by the China Grant for Talent Scientific Start-Up Project and by Natural Science Foundation of China (NSFC) under grant No. 12175134.
This work was supported by JSPS KAKENHI Grant Numbers JP22J00537 (M.S.).

\appendix

\section{Characters of SM and SU(5) groups}
\label{sec:Characters of SM group and SU(5)}

In order to see that the SM fermions are apparently unified into the $\mathrm{SU}(5)$ multiplets,
it is convenient to use the 
character of the gauge groups $G$
defined by the trace of 
a representation matrix, $R(g)$ 
($g\in G$)
\begin{align}
 \chi_R(g):= \tr[R(g)] = R(g)_{ii}\ .
\end{align}
In this paper, we use the left-handed Weyl fermions, and the representation matrices are defined for those left-handed Weyl fermions.
In this definition, the chiral nature of the representations of the fermions is encoded in the following quantity,
\begin{align}
    \Delta \chi_{R}(g) := \chi_{R}(g) - \chi_{R^{\dagger}}(g) \ ,
\end{align}
to which only the chiral fermions give non-zero contributions~\cite{GoodMan:1985bw}.

Now, let us consider $\Delta \chi_{R}$ 
in the SM model.
Since the SM model fermions are the chiral fermions, all of them contributes to $\Delta \chi_{R}$ which amounts to
\begin{align}
\label{eq:tr1}
    A_{\mathrm{SM}}(g_{\mathrm{SM}}) 
    := n_{\mathrm{gen}} \times  \sum_{i = L,\overline{d},Q,\overline{u},\overline{e}} \Delta \chi_{i}(g_{\mathrm{SM}})  \ ,
\end{align}
where $n_\mathrm{gen}=3$ is the number of the fermion generations.
As a surprising feature of the SM fermions, $A_{\mathrm{SM}}(g_{\mathrm{SM }})$ coincides with the $\Delta \chi_R$
of the $\overline{\mathbf{5}}\oplus\mathbf{10}$ representation of $\mathrm{SU}(5)$,
that is
\begin{align}
\label{eq:deltaChi GUT}
    A_{\mathrm{SM}}(g_{\mathrm{SM}}) 
    = n_{\mathrm{gen}} \times [\Delta \chi_{\overline{\mathbf{5}}}(g_{\mathrm{SM}}) +\Delta \chi_{\mathbf{10}}(g_{\mathrm{SM}})] \ ,
\end{align}
where $g_{\mathrm{SM}}$ is the $\mathrm{SU}(5)$ elements restricted to the SM gauge group.
By remembering the orthogonality and the completeness of the characters, 
Eq.\,\eqref{eq:deltaChi GUT} means that the SM fermions can be exactly embedded into the three copies of $\overline{\mathbf{5}}\oplus\mathbf{10}$.
This amazing feature cannot be explained within the SM, 
because, in general, chiral fermions consistent with SM gauge symmetry do not necessarily satisfy this property.

The fake GUT conditions in Sec.~\ref{sec:FAKE_GUT} guarantee the relation in Eq.\,\eqref{eq:deltaChi GUT} automatically.
Let us emphasize again that the quarks and leptons are not required to reside in the same $\mathrm{SU}(5)$ multiplets in the fake GUT model unlike the conventional GUT model. 
Once Eq.\,\eqref{eq:deltaChi GUT} is guaranteed by the fake GUT model, it uniquely determines the gauge charges 
of the low energy fermions under $G_{\mathrm{SM}}$ due to the orthogonality and the completeness of the characters.  
In this way, the fake GUT model 
explains why the SM fermions form the apparently complete $\mathrm{SU}(5)$ multiplets.

In the following, we list the characters of the SM and $\mathrm{SU}(5)$.
Here, we parameterize the Cartan subgroups of $G_{\mathrm{SM}}$ 
with four parameters,
$\theta_{\mathrm{U}(1)}$,
$\theta_{\mathrm{SU}(2)}$,
$\theta_{\mathrm{SU}(3),3}$,
and $\theta_{\mathrm{SU}(3),8}$.
In this case,
the character of the SM leptons and quarks are computed as follows:
\begin{align}
\chi_{L} (g_{\mathrm{SM}})
&= 
\mathrm{Tr} \left[ \mathrm{exp} \left(i \theta_{\mathrm{SU}(2)} \frac{\sigma^{3}}{2} \right) \right]
\times 
\mathrm{Tr} \left[ \mathrm{exp} \left(i \theta_{\mathrm{U}(1)} \left( -\frac{1}{2} \right) \right) \right] \nonumber \\
&= \left[ \mathrm{exp} \left( \frac{i \theta_{\mathrm{SU}(2)}}{2} \right) + \mathrm{exp} \left( - \frac{i \theta_{\mathrm{SU}(2)}}{2} \right) \right] 
\times  
\mathrm{exp} \left( - \frac{i \theta_{\mathrm{U}(1)}}{2} \right), \\
\chi_{\overline{d}} (g_{\mathrm{SM}})
&= 
\mathrm{Tr} \left[ \mathrm{exp} \left(-i \theta_{\mathrm{SU}(3),3} \frac{\lambda^{3}}{2} -i \theta_{\mathrm{SU}(3),8} \frac{\lambda^{8}}{2} \right) \right]
\times 
\mathrm{Tr} \left[ \mathrm{exp} \left(i \theta_{\mathrm{U}(1)} \frac{1}{3} \right) \right] \nonumber \\
&=
\left[ \mathrm{exp} \left( \frac{i \theta_{\mathrm{SU}(3),8}}{\sqrt{3}} \right) 
+ \mathrm{exp} \left( \frac{i \theta_{\mathrm{SU}(3),3}}{2} - \frac{ \sqrt{3} i \theta_{\mathrm{SU}(3),8}}{6} \right) 
+ \mathrm{exp} \left( - \frac{i \theta_{\mathrm{SU}(3),3}}{2} - \frac{ \sqrt{3} i \theta_{\mathrm{SU}(3),8}}{6} \right) \right] \nonumber \\
&\times  
\mathrm{exp} \left( \frac{i \theta_{\mathrm{U}(1)}}{3} \right), \\
\chi_{Q} (g_{\mathrm{SM}})
&= 
\mathrm{Tr} \left[ \mathrm{exp} \left(i \theta_{\mathrm{SU}(3),3} \frac{\lambda^{3}}{2} +i \theta_{\mathrm{SU}(3),8} \frac{\lambda^{8}}{2} \right) \right]
\times 
\mathrm{Tr} \left[ \mathrm{exp} \left(i \theta_{\mathrm{SU}(2)} \frac{\sigma^{3}}{2} \right) \right]
\times 
\mathrm{Tr} \left[ \mathrm{exp} \left(i \theta_{\mathrm{U}(1)} \frac{1}{6} \right) \right] \nonumber \\
&=
\left[ \mathrm{exp} \left( \frac{ - i \theta_{\mathrm{SU}(3),8}}{\sqrt{3}} \right) 
+ \mathrm{exp} \left( \frac{- i \theta_{\mathrm{SU}(3),3}}{2} + \frac{ \sqrt{3} i \theta_{\mathrm{SU}(3),8}}{6} \right) 
+ \mathrm{exp} \left( \frac{i \theta_{\mathrm{SU}(3),3}}{2} + \frac{ \sqrt{3} i \theta_{\mathrm{SU}(3),8}}{6} \right) \right] \nonumber \\
&\times 
\left[ \mathrm{exp} \left( \frac{i \theta_{\mathrm{SU}(2)}}{2} \right) + \mathrm{exp} \left( - \frac{i \theta_{\mathrm{SU}(2)}}{2} \right) \right] 
\times  
\mathrm{exp} \left( \frac{i \theta_{\mathrm{U}(1)}}{6} \right), \\
\chi_{\overline{u}} (g_{\mathrm{SM}})
&= 
\mathrm{Tr} \left[ \mathrm{exp} \left(-i \theta_{\mathrm{SU}(3),3} \frac{\lambda^{3}}{2} -i \theta_{\mathrm{SU}(3),8} \frac{\lambda^{8}}{2} \right) \right]
\times 
\mathrm{Tr} \left[ \mathrm{exp} \left(i \theta_{\mathrm{U}(1)} \left( -\frac{2}{3} \right) \right) \right] \nonumber \\
&= 
\left[ \mathrm{exp} \left( \frac{ i \theta_{\mathrm{SU}(3),8}}{\sqrt{3}} \right) 
+ \mathrm{exp} \left( \frac{i \theta_{\mathrm{SU}(3),3}}{2} - \frac{ \sqrt{3} i \theta_{\mathrm{SU}(3),8}}{6} \right) 
+ \mathrm{exp} \left( - \frac{i \theta_{\mathrm{SU}(3),3}}{2} - \frac{ \sqrt{3} i \theta_{\mathrm{SU}(3),8}}{6} \right) \right] \nonumber \\
&\times  
\mathrm{exp} \left( - \frac{2i \theta_{\mathrm{U}(1)}}{3} \right), \\
\chi_{\overline{e}} (g_{\mathrm{SM}})
&= 
\mathrm{Tr} \left[ \mathrm{exp} \left(i \theta_{\mathrm{U}(1)} (1) \right) \right]
= \mathrm{exp} \left( i \theta_{\mathrm{U}(1)} \right).
\end{align}

Next, let us compute the characters of the $\overline{\mathbf{5}}$ and $\mathbf{10}$.
Here, we parameterize the Cartan subgroups of $\mathrm{SU}(5)$ with four parameters, $\theta_3$, $\theta_8$, $\theta_{23}$, $\theta_{24}$.
At this time, the diagonal generators of $\overline{\mathbf{5}}$ and $\mathbf{10}$ representations correponding each parameters are the following forms:
\begin{align}
T^{3}_{\overline{\mathbf{5}}} 
&= - \frac{1}{2} \, \mathrm{diag} \left( 1, - 1, 0, 0, 0 \right), \\
T^{8}_{\overline{\mathbf{5}}} 
&= - \frac{1}{2 \sqrt{3}} \, \mathrm{diag} \left( 1, 1, -2, 0, 0 \right), \\
T^{23}_{\overline{\mathbf{5}}} 
&= - \frac{1}{2} \, \mathrm{diag} \left( 0, 0, 0, 1, -1 \right), \\
T^{24}_{\overline{\mathbf{5}}} 
&= - \frac{1}{2 \sqrt{15}} \, \mathrm{diag} \left( -2, -2, -2, 3, 3 \right), \\
T^{3}_{\mathbf{10}} 
&= \frac{1}{2} \, \mathrm{diag} \left( 1, - 1, 0, 1, - 1, 0, - 1, 1, 0, 0 \right), \\
T^{8}_{\mathbf{10}} 
&= \frac{1}{2 \sqrt{3}} \, \mathrm{diag} \left( 1, 1, -2, 1, 1, -2, -1, -1, 2, 0 \right), \\
T^{23}_{\mathbf{10}} 
&= \frac{1}{2} \, \mathrm{diag} \left( 1, 1, 1, -1, -1, -1, 0, 0, 0, 0 \right), \\
T^{24}_{\mathbf{10}} 
&= \frac{1}{2 \sqrt{15}} \, \mathrm{diag} \left( 1, 1, 1, 1, 1, 1, -4, -4, -4, 6 \right).
\end{align}
In this case,
$\chi_{\overline{\mathbf{5}}}$ and $\chi_{\mathbf{10}}$ are
\begin{align}
\chi_{\overline{\mathbf{5}}} (g_{\mathrm{SM}})
&= 
\mathrm{Tr} \left[ \mathrm{exp} 
\left(i \theta_{3} T^{3}_{\overline{\mathbf{5}}} 
+ i \theta_{8} T^{8}_{\overline{\mathbf{5}}} 
+ i \theta_{23} T^{23}_{\overline{\mathbf{5}}} 
+ i \theta_{24} \sqrt{\frac{5}{3}} T^{24}_{\overline{\mathbf{5}}}  \right) \right] \nonumber \\
&= 
\left[ \mathrm{exp} \left( \frac{i \theta_{23}}{2} \right) + \mathrm{exp} \left( - \frac{i \theta_{23}}{2} \right) \right] 
\times  
\mathrm{exp} \left( - \frac{i \theta_{24}}{2} \right) \nonumber \\
&+
\left[ \mathrm{exp} \left( \frac{i \theta_{8}}{\sqrt{3}} \right) 
+ \mathrm{exp} \left( \frac{i \theta_{3}}{2} - \frac{ \sqrt{3} i \theta_{8}}{6} \right) 
+ \mathrm{exp} \left( - \frac{i \theta_{3}}{2} - \frac{ \sqrt{3} i \theta_{8}}{6} \right) \right] \nonumber \\
&\times  
\mathrm{exp} \left( \frac{i \theta_{24}}{3} \right), \\
\chi_{\mathbf{10}} (g_{\mathrm{SM}})
&= 
\mathrm{Tr} \left[ \mathrm{exp} 
\left(i \theta_{3} T^{3}_{\mathbf{10}} 
+ i \theta_{8} T^{8}_{\mathbf{10}} 
+ i \theta_{23} T^{23}_{\mathbf{10}} 
+ i \theta_{24} \sqrt{\frac{5}{3}} T^{24}_{\mathbf{10}}  \right) \right] \nonumber \\
&= \left[ \mathrm{exp} \left( \frac{ - i \theta_{8}}{\sqrt{3}} \right) 
+ \mathrm{exp} \left( \frac{- i \theta_{3}}{2} + \frac{ \sqrt{3} i \theta_{8}}{6} \right) 
+ \mathrm{exp} \left( \frac{i \theta_{3}}{2} + \frac{ \sqrt{3} i \theta_{8}}{6} \right) \right] \nonumber \\
&\times 
\left[ \mathrm{exp} \left( \frac{i \theta_{23}}{2} \right) + \mathrm{exp} \left( - \frac{i \theta_{23}}{2} \right) \right] 
\times  
\mathrm{exp} \left( \frac{i \theta_{24}}{6} \right) \nonumber \\
&+ 
\left[ \mathrm{exp} \left( \frac{ i \theta_{8}}{\sqrt{3}} \right) 
+ \mathrm{exp} \left( \frac{i \theta_{3}}{2} - \frac{ \sqrt{3} i \theta_{8}}{6} \right) 
+ \mathrm{exp} \left( - \frac{i \theta_{3}}{2} - \frac{ \sqrt{3} i \theta_{8}}{6} \right) \right] \nonumber \\
&\times  
\mathrm{exp} \left( - \frac{2i \theta_{24}}{3} \right) \nonumber \\
&+ 
\mathrm{exp} \left( i \theta_{24} \right).
\end{align}
If we identify $\theta_3$, $\theta_8$, $\theta_{23}$ and $\theta_{24}$ with $\theta_{\mathrm{SU}(3),3}$, $\theta_{\mathrm{SU}(3),8}$, $\theta_{\mathrm{SU}(2)}$ and $\theta_{\mathrm{U}(1)}$ respectively,
we can find that $\chi_{\overline{\mathbf{5}}} (g_{\mathrm{SM}})$ and $\chi_{\mathbf{10}} (g_{\mathrm{SM}})$ correspond to $\chi_{L} (g_{\mathrm{SM}})
+ \chi_{\overline{d}} (g_{\mathrm{SM}})$ and
$\chi_{Q} (g_{\mathrm{SM}})
+ \chi_{\overline{u}} (g_{\mathrm{SM}})
+ \chi_{\overline{e}} (g_{\mathrm{SM}})$ respectively.

\section{RG Equations}
\label{sec:RGE}
We show the RG equations of the gauge couplings of the $\mathrm{SU}(5) \times \mathrm{U}(2)_H$ model and $\mathrm{SU}(5) \times \mathrm{SU}(3)_H$ model up to two-loop level.
We used the program PyR@TE 3 \cite{Sartore:2020gou} for the calculation.
Here, we neglect the Yukawa and Higgs couplings.
\subsubsection*{$\mathrm{SU}(5) \times \mathrm{U}(2)_H$ model}
The matter fields are given in Tab.\,\ref{tab:U2particlecontents}.
\begin{align}
  (4 \pi)^2  \frac{d g_5}{ d \log \mu} &=- \frac{83}{6} g_{5}^{3} + \frac{1}{(4\pi)^2} \left( -  \frac{373}{3} g_{5}^{5}
+ 3 g_{5}^{3} g_{2H}^{2}
+ g_{5}^{3} g_{1H}^{2} \right),\\
  (4 \pi)^2  \frac{d g_{2H}}{ d \log \mu} &=- \frac{13}{3} g_{2H}^{3}+ \frac{1}{(4\pi)^2} \left(  24 g_{5}^{2} g_{2H}^{3}
-  \frac{47}{6} g_{2H}^{5}
+ \frac{9}{2} g_{2H}^{3} g_{1H}^{2}
\right),\\
  (4 \pi)^2  \frac{d g_{1H}}{ d \log \mu} &=7 g_{1H}^{3}+ \frac{1}{(4\pi)^2} \left(  24 g_{5}^{2} + 24 g_{5}^{2} g_{1H}^{3}
+ \frac{27}{2} g_{2H}^{2} g_{1H}^{3}
+ \frac{33}{2} g_{1H}^{5}
\right).
\end{align}
\subsubsection*{$\mathrm{SU}(5) \times \mathrm{SU}(3)_H$ model}
The matter fields are given in Tab.\,\ref{tab:SU3particlecontents}.
\begin{align}
  (4 \pi)^2  \frac{d g_5}{ d \log \mu} &=- \frac{41}{3} g_{5}^{3} + \frac{1}{(4\pi)^2} \left(-  \frac{1768}{15} g_{5}^{5}
+ 8 g_{3H}^{2} g_{5}^{3} \right),\\
  (4 \pi)^2  \frac{d g_{3H}}{ d \log \mu} &=-  \frac{15}{2} g_{3H}^{3}
+ \frac{1}{(4\pi)^2} \left(  
 24 g_{3H}^{3} g_{5}^{2}
- 21 g_{3H}^{5}
\right).
\end{align}

\section{On global lepton and baryon symmetry}
\label{sec:On global lepton and baryon symmetry}

As discussed in subsection~\ref{sec:Global Symmetry},
the global lepton symmetry plays a crucial role to make the $\mathrm{SU}(5)\times \mathrm{U}(2)_H$ model phenomenologically viable.
The global lepton symmetry is more important for $\mathrm{SU}(5)\times \mathrm{SU}(3)_H$ model due to the smaller $X$ gauge boson mass.
In this appendix, we give an example of the model in which the high quality lepton symmetry originates from a 
discrete gauge symmetry in $\mathrm{SU}(5)\times \mathrm{SU}(3)_H$ model.

Before discussing the lepton symmetry, 
we first provide 
a concrete model to generate the higher dimensional operators used in Eqs.\,\eqref{eq:su(3) mass}
and \eqref{eq:su3_Yukawa}.
As for the  mixing term in Eq.\,\eqref{eq:su(3) mass},
we consider a complex scalar
$\phi_3'$ in $(\overline{\mathbf{10}},{\mathbf{3}})$ 
representation which has a Yukawa coupling 
$\overline{L}_T\phi_3' \mathbf{10}$
trilinear coupling $\phi_3 \phi_3' \phi_3$.
When the mass of $\phi_3'$ is larger than 
the fake GUT scale while the coefficient of the trilinear coupling is of order of the fake GUT scale, the VEV of $\phi_3'$ is aligned to
the $\phi_3' \sim \phi_3^\dagger \phi_3^\dagger$.
In this way, the higher dimensional lepton mixing mass is achieved by
\begin{align}
\label{eq:mixingUV}
    \mathcal{L} = \lambda_{10,ij} \overline{L}_T \phi'_3 \mathbf{10}_{j} \to 
    \mathcal{L}_\mathrm{eff} \sim \frac{\mu_{\phi\phi'\phi}}{M_{\phi'}^2}\lambda_{10,ij} \overline{L}_T \phi_3^\dagger
    \phi_3^\dagger \mathbf{10}_{j} \ .
\end{align}
Here, $M_{\phi'}$ and $\mu_{\phi\phi'\phi}$
are the mass and the coefficient of the trilinear coupling of $\phi'$, respectively.

To generate the origin of the Yukawa interactions in Eq.\,\eqref{eq:su3_Yukawa}, we introduce other Dirac fermions, $(H_F,\overline{H}_F)$ with the Yukawa interactions,
\begin{align}
\label{eq:YukawaUV}
\mathcal{L}_{YL} 
= - M_{H \overline{H}} H_{F} \overline{H}_{F} + a_{i} L_{Ti} A \overline{H}_{F} + b_{j} H_{F} L_{Tj} H_{3} 
+ h.c..
\end{align}
We again assume that ($H_F$,$\overline{H}_F$) is heavier than the fake GUT scale.
By integrating out $H_F$, $\overline{H}_F$,
we obtain, 
\begin{align}
\mathcal{L}_{\mathrm{eff}} 
= \frac{a_{i} b_{j}}{M_{H \overline{H}}}  {L}_{Ti}^{\alpha} A_{\alpha}^{\beta} {L}_{Tj}^{\gamma} H_{3}^{\delta} \epsilon_{\beta \gamma \delta}
+ h.c..
\end{align}
Since we have the other Yukawa coupling in Eq.\,\eqref{eq:SU3Yukawa}, one pair of $(H_F,\overline{H}_F)$ 
can reproduce the masses of the SM leptons
with appropriate choice of $a_i$, $b_i$
and $(Y_{LT})_{ij}$.

In Sec.\,\ref{sec:Global Symmetry}
we assumed a global lepton symmetry
to suppress the proton decay rate.
In the $\mathrm{SU}(5)\times \mathrm{SU}(3)_H$
model, the lepton symmetry is required to suppress
$m_T$ and $\lambda_T$. 
In the present model, however, the 
lepton Yukawa interactions in
Eq.\,\eqref{eq:SU3Yukawa} and  \eqref{eq:su3_Yukawa} include $L_T^2$,
and hence,
the lepton symmetry can not be the continuous $\mathrm{U}(1)$ symmetry.
Instead, we consider a discrete $\mathbb{Z}_{2n}$ $(n\in \mathbb{N})$ symmetry, under which $L_T$'s have charge $n$.
The discrete $\mathbb{Z}_{2n}$ 
symmetry forbids only the unwanted $m_T$
and $\lambda_T$ terms when
$(H_F,\overline{H}_F)$ also have $\mathbb{Z}_{2n}$ charge $n$ 
while other fields are neutral under $\mathbb{Z}_{2n}$ (see Tab.~\ref{tab:z4 particle}).

As mentioned earlier, it is argued that all global symmetries are broken by quantum gravity effects\,(see e.g., Refs.\,\cite{Hawking:1987mz,Lavrelashvili:1987jg,Giddings:1988cx,Coleman:1988tj,Gilbert:1989nq,Kallosh:1995hi,Banks:2010zn}).
Thus, we seek a possibility to realize $\mathbb{Z}_{2n}$ symmetry. 
The anomaly coefficients of the gauged discrete symmetry come only 
from $L_T(n)$
since $(H_F,\overline{H}_F)$ contributions trivially cancels;
 \begin{align}
    &\mathbb{Z}_{2n}\times [\mathrm{SU}(3)_{H}]^{2}  = n \times 3  \equiv n\,\,  {(\mathrm{mod} \,2n)}\ , \\
    &\mathbb{Z}_{2n}\times [\mathrm{gravity}]^{2} = n \times 3 \times 3  \equiv 0\,\, {(\mathrm{mod} \,n)}\ , \\
    &\mathbb{Z}_{2n}^3 = n^{3} \times 3 \times 3  \equiv 0 \,\,{(\mathrm{mod}\,\,n^{3} \,\mathrm{or}~2n)}\ , \label{eq:Z3cubic}
 \end{align}
which should be vanishing for 
the $\mathbb{Z}_{2n}$ symmetry to be a gauge symmetry~\cite{Ibanez:1991hv,Ibanez:1991pr,Ibanez:1992ji,Csaki:1997aw}.
The third condition in Eq.\,\ref{eq:Z3cubic}
can be always satisfied by choosing the normalization of $n$ appropriately~\cite{Banks:1991xj}.
Thus, it does not lead to useful constraints 
on the particle contents.
Here, the first, second and third quantities in the each  multiplication represent $\mathbb{Z}_{2n}$ charge, the number of particles with same charges and degrees of freedom of $\mathrm{SU}(3)_H$.
Thus, to realize the gauged $\mathbb{Z}_{2n}$ symmetry, we need additional $\mathrm{SU}(3)_H$ charged fermions which are chiral under $\mathbb{Z}_{2n}$.

To cancel the anomaly, $\mathbb{Z}_{2n}\times [\mathrm{SU}(3)_{H}]^{2}$, we introduce 
a pair of ($\Theta$, $\overline{\Theta}$)
which are (anti)-fundamental representation of 
$\mathrm{SU}(3)_H$ and have  $\mathbb{Z}_{2n}$ charges
($n$,$0$).
In this case, all the anomaly coefficients 
vanish, and $\mathbb{Z}_{2n}$ can be the gauge symmetry.
However,  $\mathbb{Z}_{2n}$ must 
be broken to give a mass to the pair, ($\Theta$, $\overline{\Theta}$), by the VEV of a complex scalar $\phi_Z$ with the $\mathbb{Z}_{2n}$ charge $n$.
Such a VEV of $\phi_Z$ also generates $m_T$,
which results in the large lepton mixing angles.

To avoid this problem, we assume that the 
additional chiral fermions have smaller $\mathbb{Z}_{2n}$ charges.
Concretely, we consider $n=4$, and hence, $\mathbb{Z}_{8}$ symmetry, with the 
$\mathbb{Z}_{8}$ charges of ($\Theta$, $\overline{\Theta}$) being 
($2$,$0$) (see Tab.\,\ref{tab:z4 particle}).
In this case, the anomaly coefficients of $\mathbb{Z}_{8}\times [\mathrm{SU}(3)_{H}]^{2}$
and $\mathbb{Z}_{8}\times [\mathrm{gravity}]^{2}$ require two pairs of 
($\Theta$, $\overline{\Theta}$).
In this way, we achieve the anomaly-free $\mathbb{Z}_8$ symmetry.

The masses of 
the pairs of ($\Theta$, $\overline{\Theta}$) 
are given by 
the VEV of the complex field $\phi_Z$ with 
a $\mathbb{Z}_8$ charge $6$, $v_Z = \langle \phi_Z \rangle$.
The mass of a pair ($H_F$, $\overline{H}_F$) 
is not forbidden by the $\mathbb{Z}_8$ symmetry.%
\footnote{The fields $L_T$ and $H_F$ have the same gauge charges, and hence, they mix with each other.
Such mixing does not alter our conclusions.}
The lepton symmetry breaking parameters 
$m_T$ and $\lambda_{TAT}$ are, on the other hand, suppressed by the cutoff scale which is now taken to be the Planck scale, 
$M_\mathrm{Pl}$,
\begin{align}
    m_T \sim \frac{v_Z^2}{M_\mathrm{Pl}} \ ,
    \quad \lambda_{TAT} v_A \sim \frac{v_Z^2}{M_\mathrm{Pl}^2}v_A \ .
\end{align}
To achieve those mass parameters smaller than $10^{-4}$\,GeV, we find that the breaking scale of $\mathbb{Z}_8$ is limited from above,
\begin{align}
\label{eq:vZlimit}
   v_Z < \order{10^7}\,\mathrm{GeV}\ .
\end{align}
In summary,
the extra particles ($\Theta$, $\overline{\Theta}$)
 have masses of $v_Z \lesssim 10^7$\,GeV, while the lepton symmetry breaking mass parameter $m_T$ is highly suppressed.

\begin{table}[tb]
 \begin{center}
  \caption{Field contents of fermions and a scalar with the $\mathbb{Z}_{2n}$ charge and pairing fermions for $n=4$.
  SM lepton doublets are in $L_{T}$'s.
  The $\mathbb{Z}_8$ symmetry is free from anomaly including $\mathbb{Z}_8^3$.
  In practice, this $\mathbb{Z}_8$ symmetry is equivalent to $\mathbb{Z}_4$ symmetry.}
  \begin{tabular}{|c||c|c|c|}  \hline
     & $\mathrm{SU}(5)$ & $\mathrm{SU}(3)_{H}$ & $\mathbb{Z}_{8}$ \\ \hline \hline
    $L_{T1,2,3}$ & \bf{1} & \bf{3} & $4$ \\ 
    $\overline{L}_{T1,2,3}$ & \bf{1} & $\overline{\bf{3}}$ & $0$ \\ 
    $H_{F}$ & \bf{1} & \bf{3} & $4$ \\ 
    $\overline{H}_{F}$ & \bf{1} & $\overline{\bf{3}}$ & $4$ \\
    $\Theta_{1,2}$ & \bf{1} & \bf{3} & $2$ \\
    $\overline{\Theta}_{1,2}$ & \bf{1} & $\overline{\bf{3}}$ & $0$ \\ 
    \hline\hline
    $\phi_{Z}$ & \bf{1} & \bf{1} & $6$ \\ \hline
  \end{tabular}
  \label{tab:z4 particle}
 \end{center} 
\end{table}

When a discrete symmetry is broken spontaneously, 
the domain walls are formed.
To avoid the cosmological domain wall problem,
we need to assume that at least the reheating temperature after inflation should be lower than the $\mathbb{Z}_8$ breaking scale, $v_Z$.
Combined with the upper limit on $v_Z$ in Eq.\,\eqref{eq:vZlimit}, 
the upper limit on the reheating temperature is given by $T_{\mathrm{rh}} \lesssim \mathcal{O}(10^{7}) \, \mathrm{GeV} $.

The running of the gauge coupling of the $\mathrm{SU}(3)_{H}$ is modified because of the addition of multiple Dirac fermions and scalar particles in this example model.
However, the asymptotic freedom $\mathrm{SU}(3)_{H}$ gauge interaction is still preserved in the present setup.

\bibliographystyle{apsrev4-1}
\bibliography{bibtex}

\begin{thebibliography}{56}%
\makeatletter
\providecommand \@ifxundefined [1]{%
 \@ifx{#1\undefined}
}%
\providecommand \@ifnum [1]{%
 \ifnum #1\expandafter \@firstoftwo
 \else \expandafter \@secondoftwo
 \fi
}%
\providecommand \@ifx [1]{%
 \ifx #1\expandafter \@firstoftwo
 \else \expandafter \@secondoftwo
 \fi
}%
\providecommand \natexlab [1]{#1}%
\providecommand \enquote  [1]{``#1''}%
\providecommand \bibnamefont  [1]{#1}%
\providecommand \bibfnamefont [1]{#1}%
\providecommand \citenamefont [1]{#1}%
\providecommand \href@noop [0]{\@secondoftwo}%
\providecommand \href [0]{\begingroup \@sanitize@url \@href}%
\providecommand \@href[1]{\@@startlink{#1}\@@href}%
\providecommand \@@href[1]{\endgroup#1\@@endlink}%
\providecommand \@sanitize@url [0]{\catcode `\\12\catcode `\$12\catcode
  `\&12\catcode `\#12\catcode `\^12\catcode `\_12\catcode `\%12\relax}%
\providecommand \@@startlink[1]{}%
\providecommand \@@endlink[0]{}%
\providecommand \url  [0]{\begingroup\@sanitize@url \@url }%
\providecommand \@url [1]{\endgroup\@href {#1}{\urlprefix }}%
\providecommand \urlprefix  [0]{URL }%
\providecommand \Eprint [0]{\href }%
\providecommand \doibase [0]{http://dx.doi.org/}%
\providecommand \selectlanguage [0]{\@gobble}%
\providecommand \bibinfo  [0]{\@secondoftwo}%
\providecommand \bibfield  [0]{\@secondoftwo}%
\providecommand \translation [1]{[#1]}%
\providecommand \BibitemOpen [0]{}%
\providecommand \bibitemStop [0]{}%
\providecommand \bibitemNoStop [0]{.\EOS\space}%
\providecommand \EOS [0]{\spacefactor3000\relax}%
\providecommand \BibitemShut  [1]{\csname bibitem#1\endcsname}%
\let\auto@bib@innerbib\@empty
\bibitem [{\citenamefont {Foot}\ \emph {et~al.}(1989)\citenamefont {Foot},
  \citenamefont {Lew}, \citenamefont {Volkas},\ and\ \citenamefont
  {Joshi}}]{Foot:1988qx}%
  \BibitemOpen
  \bibfield  {author} {\bibinfo {author} {\bibfnamefont {R.}~\bibnamefont
  {Foot}}, \bibinfo {author} {\bibfnamefont {H.}~\bibnamefont {Lew}}, \bibinfo
  {author} {\bibfnamefont {R.~R.}\ \bibnamefont {Volkas}}, \ and\ \bibinfo
  {author} {\bibfnamefont {G.~C.}\ \bibnamefont {Joshi}},\ }\href {\doibase
  10.1103/PhysRevD.39.3411} {\bibfield  {journal} {\bibinfo  {journal} {Phys.
  Rev. D}\ }\textbf {\bibinfo {volume} {39}},\ \bibinfo {pages} {3411}
  (\bibinfo {year} {1989})}\BibitemShut {NoStop}%
\bibitem [{\citenamefont {Knochel}\ and\ \citenamefont
  {Wetterich}(2012)}]{Knochel:2011ng}%
  \BibitemOpen
  \bibfield  {author} {\bibinfo {author} {\bibfnamefont {A.}~\bibnamefont
  {Knochel}}\ and\ \bibinfo {author} {\bibfnamefont {C.}~\bibnamefont
  {Wetterich}},\ }\href {\doibase 10.1016/j.physletb.2011.09.004} {\bibfield
  {journal} {\bibinfo  {journal} {Phys. Lett. B}\ }\textbf {\bibinfo {volume}
  {706}},\ \bibinfo {pages} {320} (\bibinfo {year} {2012})},\ \Eprint
  {http://arxiv.org/abs/1106.2609} {arXiv:1106.2609 [hep-ph]} \BibitemShut
  {NoStop}%
\bibitem [{\citenamefont {Cebola}\ \emph {et~al.}(2014)\citenamefont {Cebola},
  \citenamefont {Emmanuel-Costa}, \citenamefont {Gonz\'alez~Felipe},\ and\
  \citenamefont {Sim\~oes}}]{Cebola:2014qfa}%
  \BibitemOpen
  \bibfield  {author} {\bibinfo {author} {\bibfnamefont {L.~M.}\ \bibnamefont
  {Cebola}}, \bibinfo {author} {\bibfnamefont {D.}~\bibnamefont
  {Emmanuel-Costa}}, \bibinfo {author} {\bibfnamefont {R.}~\bibnamefont
  {Gonz\'alez~Felipe}}, \ and\ \bibinfo {author} {\bibfnamefont
  {C.}~\bibnamefont {Sim\~oes}},\ }\href {\doibase 10.1103/PhysRevD.90.125037}
  {\bibfield  {journal} {\bibinfo  {journal} {Phys. Rev. D}\ }\textbf {\bibinfo
  {volume} {90}},\ \bibinfo {pages} {125037} (\bibinfo {year} {2014})},\
  \Eprint {http://arxiv.org/abs/1409.0805} {arXiv:1409.0805 [hep-ph]}
  \BibitemShut {NoStop}%
\bibitem [{\citenamefont {Georgi}\ and\ \citenamefont
  {Glashow}(1974)}]{Georgi:1974sy}%
  \BibitemOpen
  \bibfield  {author} {\bibinfo {author} {\bibfnamefont {H.}~\bibnamefont
  {Georgi}}\ and\ \bibinfo {author} {\bibfnamefont {S.~L.}\ \bibnamefont
  {Glashow}},\ }\href {\doibase 10.1103/PhysRevLett.32.438} {\bibfield
  {journal} {\bibinfo  {journal} {Phys. Rev. Lett.}\ }\textbf {\bibinfo
  {volume} {32}},\ \bibinfo {pages} {438} (\bibinfo {year} {1974})}\BibitemShut
  {NoStop}%
\bibitem [{\citenamefont {Georgi}\ \emph {et~al.}(1974)\citenamefont {Georgi},
  \citenamefont {Quinn},\ and\ \citenamefont {Weinberg}}]{Georgi:1974yf}%
  \BibitemOpen
  \bibfield  {author} {\bibinfo {author} {\bibfnamefont {H.}~\bibnamefont
  {Georgi}}, \bibinfo {author} {\bibfnamefont {H.~R.}\ \bibnamefont {Quinn}}, \
  and\ \bibinfo {author} {\bibfnamefont {S.}~\bibnamefont {Weinberg}},\ }\href
  {\doibase 10.1103/PhysRevLett.33.451} {\bibfield  {journal} {\bibinfo
  {journal} {Phys. Rev. Lett.}\ }\textbf {\bibinfo {volume} {33}},\ \bibinfo
  {pages} {451} (\bibinfo {year} {1974})}\BibitemShut {NoStop}%
\bibitem [{\citenamefont {Buras}\ \emph {et~al.}(1978)\citenamefont {Buras},
  \citenamefont {Ellis}, \citenamefont {Gaillard},\ and\ \citenamefont
  {Nanopoulos}}]{Buras:1977yy}%
  \BibitemOpen
  \bibfield  {author} {\bibinfo {author} {\bibfnamefont {A.~J.}\ \bibnamefont
  {Buras}}, \bibinfo {author} {\bibfnamefont {J.~R.}\ \bibnamefont {Ellis}},
  \bibinfo {author} {\bibfnamefont {M.~K.}\ \bibnamefont {Gaillard}}, \ and\
  \bibinfo {author} {\bibfnamefont {D.~V.}\ \bibnamefont {Nanopoulos}},\ }\href
  {\doibase 10.1016/0550-3213(78)90214-6} {\bibfield  {journal} {\bibinfo
  {journal} {Nucl. Phys. B}\ }\textbf {\bibinfo {volume} {135}},\ \bibinfo
  {pages} {66} (\bibinfo {year} {1978})}\BibitemShut {NoStop}%
\bibitem [{\citenamefont {Zyla}\ \emph {et~al.}(2020)\citenamefont {Zyla} \emph
  {et~al.}}]{ParticleDataGroup:2020ssz}%
  \BibitemOpen
  \bibfield  {author} {\bibinfo {author} {\bibfnamefont {P.~A.}\ \bibnamefont
  {Zyla}} \emph {et~al.} (\bibinfo {collaboration} {Particle Data Group}),\
  }\href {\doibase 10.1093/ptep/ptaa104} {\bibfield  {journal} {\bibinfo
  {journal} {PTEP}\ }\textbf {\bibinfo {volume} {2020}},\ \bibinfo {pages}
  {083C01} (\bibinfo {year} {2020})}\BibitemShut {NoStop}%
\bibitem [{\citenamefont {Takenaka}\ \emph {et~al.}(2020)\citenamefont
  {Takenaka} \emph {et~al.}}]{Super-Kamiokande:2020wjk}%
  \BibitemOpen
  \bibfield  {author} {\bibinfo {author} {\bibfnamefont {A.}~\bibnamefont
  {Takenaka}} \emph {et~al.} (\bibinfo {collaboration} {Super-Kamiokande}),\
  }\href {\doibase 10.1103/PhysRevD.102.112011} {\bibfield  {journal} {\bibinfo
   {journal} {Phys. Rev. D}\ }\textbf {\bibinfo {volume} {102}},\ \bibinfo
  {pages} {112011} (\bibinfo {year} {2020})},\ \Eprint
  {http://arxiv.org/abs/2010.16098} {arXiv:2010.16098 [hep-ex]} \BibitemShut
  {NoStop}%
\bibitem [{\citenamefont {Abe}\ \emph {et~al.}(2018)\citenamefont {Abe} \emph
  {et~al.}}]{Hyper-Kamiokande:2018ofw}%
  \BibitemOpen
  \bibfield  {author} {\bibinfo {author} {\bibfnamefont {K.}~\bibnamefont
  {Abe}} \emph {et~al.} (\bibinfo {collaboration} {Hyper-Kamiokande}),\
  }\href@noop {} {\  (\bibinfo {year} {2018})},\ \Eprint
  {http://arxiv.org/abs/1805.04163} {arXiv:1805.04163 [physics.ins-det]}
  \BibitemShut {NoStop}%
\bibitem [{\citenamefont {An}\ \emph {et~al.}(2016)\citenamefont {An} \emph
  {et~al.}}]{JUNO:2015zny}%
  \BibitemOpen
  \bibfield  {author} {\bibinfo {author} {\bibfnamefont {F.}~\bibnamefont {An}}
  \emph {et~al.} (\bibinfo {collaboration} {JUNO}),\ }\href {\doibase
  10.1088/0954-3899/43/3/030401} {\bibfield  {journal} {\bibinfo  {journal} {J.
  Phys. G}\ }\textbf {\bibinfo {volume} {43}},\ \bibinfo {pages} {030401}
  (\bibinfo {year} {2016})},\ \Eprint {http://arxiv.org/abs/1507.05613}
  {arXiv:1507.05613 [physics.ins-det]} \BibitemShut {NoStop}%
\bibitem [{\citenamefont {Abi}\ \emph {et~al.}(2021)\citenamefont {Abi} \emph
  {et~al.}}]{DUNE:2020fgq}%
  \BibitemOpen
  \bibfield  {author} {\bibinfo {author} {\bibfnamefont {B.}~\bibnamefont
  {Abi}} \emph {et~al.} (\bibinfo {collaboration} {DUNE}),\ }\href {\doibase
  10.1140/epjc/s10052-021-09007-w} {\bibfield  {journal} {\bibinfo  {journal}
  {Eur. Phys. J. C}\ }\textbf {\bibinfo {volume} {81}},\ \bibinfo {pages} {322}
  (\bibinfo {year} {2021})},\ \Eprint {http://arxiv.org/abs/2008.12769}
  {arXiv:2008.12769 [hep-ex]} \BibitemShut {NoStop}%
\bibitem [{\citenamefont {Ibe}\ \emph {et~al.}(2019)\citenamefont {Ibe},
  \citenamefont {Shirai}, \citenamefont {Suzuki},\ and\ \citenamefont
  {Yanagida}}]{Ibe:2019ifm}%
  \BibitemOpen
  \bibfield  {author} {\bibinfo {author} {\bibfnamefont {M.}~\bibnamefont
  {Ibe}}, \bibinfo {author} {\bibfnamefont {S.}~\bibnamefont {Shirai}},
  \bibinfo {author} {\bibfnamefont {M.}~\bibnamefont {Suzuki}}, \ and\ \bibinfo
  {author} {\bibfnamefont {T.~T.}\ \bibnamefont {Yanagida}},\ }\href {\doibase
  10.1103/PhysRevD.100.055024} {\bibfield  {journal} {\bibinfo  {journal}
  {Phys. Rev. D}\ }\textbf {\bibinfo {volume} {100}},\ \bibinfo {pages}
  {055024} (\bibinfo {year} {2019})},\ \Eprint
  {http://arxiv.org/abs/1906.02977} {arXiv:1906.02977 [hep-ph]} \BibitemShut
  {NoStop}%
\bibitem [{\citenamefont {'t~Hooft}(1980)}]{tHooft:1979rat}%
  \BibitemOpen
  \bibfield  {author} {\bibinfo {author} {\bibfnamefont {G.}~\bibnamefont
  {'t~Hooft}},\ }\href {\doibase 10.1007/978-1-4684-7571-5_9} {\bibfield
  {journal} {\bibinfo  {journal} {NATO Sci. Ser. B}\ }\textbf {\bibinfo
  {volume} {59}},\ \bibinfo {pages} {135} (\bibinfo {year} {1980})}\BibitemShut
  {NoStop}%
\bibitem [{\citenamefont {Murayama}\ \emph {et~al.}(1992)\citenamefont
  {Murayama}, \citenamefont {Okada},\ and\ \citenamefont
  {Yanagida}}]{Murayama:1991ew}%
  \BibitemOpen
  \bibfield  {author} {\bibinfo {author} {\bibfnamefont {H.}~\bibnamefont
  {Murayama}}, \bibinfo {author} {\bibfnamefont {Y.}~\bibnamefont {Okada}}, \
  and\ \bibinfo {author} {\bibfnamefont {T.}~\bibnamefont {Yanagida}},\ }\href
  {\doibase 10.1143/PTP.88.791} {\bibfield  {journal} {\bibinfo  {journal}
  {Prog. Theor. Phys.}\ }\textbf {\bibinfo {volume} {88}},\ \bibinfo {pages}
  {791} (\bibinfo {year} {1992})}\BibitemShut {NoStop}%
\bibitem [{\citenamefont {Hisano}\ \emph {et~al.}(1994)\citenamefont {Hisano},
  \citenamefont {Murayama},\ and\ \citenamefont {Yanagida}}]{Hisano:1993uk}%
  \BibitemOpen
  \bibfield  {author} {\bibinfo {author} {\bibfnamefont {J.}~\bibnamefont
  {Hisano}}, \bibinfo {author} {\bibfnamefont {H.}~\bibnamefont {Murayama}}, \
  and\ \bibinfo {author} {\bibfnamefont {T.}~\bibnamefont {Yanagida}},\ }\href
  {\doibase 10.1103/PhysRevD.49.4966} {\bibfield  {journal} {\bibinfo
  {journal} {Phys. Rev. D}\ }\textbf {\bibinfo {volume} {49}},\ \bibinfo
  {pages} {4966} (\bibinfo {year} {1994})}\BibitemShut {NoStop}%
\bibitem [{\citenamefont {Chigusa}\ and\ \citenamefont
  {Moroi}(2017)}]{Chigusa:2017drd}%
  \BibitemOpen
  \bibfield  {author} {\bibinfo {author} {\bibfnamefont {S.}~\bibnamefont
  {Chigusa}}\ and\ \bibinfo {author} {\bibfnamefont {T.}~\bibnamefont
  {Moroi}},\ }\href {\doibase 10.1093/ptep/ptx062} {\bibfield  {journal}
  {\bibinfo  {journal} {PTEP}\ }\textbf {\bibinfo {volume} {2017}},\ \bibinfo
  {pages} {063B05} (\bibinfo {year} {2017})},\ \Eprint
  {http://arxiv.org/abs/1702.00790} {arXiv:1702.00790 [hep-ph]} \BibitemShut
  {NoStop}%
\bibitem [{\citenamefont {Hall}\ and\ \citenamefont
  {Harigaya}(2018)}]{Hall:2018let}%
  \BibitemOpen
  \bibfield  {author} {\bibinfo {author} {\bibfnamefont {L.~J.}\ \bibnamefont
  {Hall}}\ and\ \bibinfo {author} {\bibfnamefont {K.}~\bibnamefont
  {Harigaya}},\ }\href {\doibase 10.1007/JHEP10(2018)130} {\bibfield  {journal}
  {\bibinfo  {journal} {JHEP}\ }\textbf {\bibinfo {volume} {10}},\ \bibinfo
  {pages} {130} (\bibinfo {year} {2018})},\ \Eprint
  {http://arxiv.org/abs/1803.08119} {arXiv:1803.08119 [hep-ph]} \BibitemShut
  {NoStop}%
\bibitem [{\citenamefont {Hall}\ and\ \citenamefont
  {Harigaya}(2019)}]{Hall:2019qwx}%
  \BibitemOpen
  \bibfield  {author} {\bibinfo {author} {\bibfnamefont {L.~J.}\ \bibnamefont
  {Hall}}\ and\ \bibinfo {author} {\bibfnamefont {K.}~\bibnamefont
  {Harigaya}},\ }\href {\doibase 10.1007/JHEP11(2019)033} {\bibfield  {journal}
  {\bibinfo  {journal} {JHEP}\ }\textbf {\bibinfo {volume} {11}},\ \bibinfo
  {pages} {033} (\bibinfo {year} {2019})},\ \Eprint
  {http://arxiv.org/abs/1905.12722} {arXiv:1905.12722 [hep-ph]} \BibitemShut
  {NoStop}%
\bibitem [{\citenamefont {Bhattacherjee}\ \emph {et~al.}(2013)\citenamefont
  {Bhattacherjee}, \citenamefont {Evans}, \citenamefont {Ibe}, \citenamefont
  {Matsumoto},\ and\ \citenamefont {Yanagida}}]{Bhattacherjee:2013gr}%
  \BibitemOpen
  \bibfield  {author} {\bibinfo {author} {\bibfnamefont {B.}~\bibnamefont
  {Bhattacherjee}}, \bibinfo {author} {\bibfnamefont {J.~L.}\ \bibnamefont
  {Evans}}, \bibinfo {author} {\bibfnamefont {M.}~\bibnamefont {Ibe}}, \bibinfo
  {author} {\bibfnamefont {S.}~\bibnamefont {Matsumoto}}, \ and\ \bibinfo
  {author} {\bibfnamefont {T.~T.}\ \bibnamefont {Yanagida}},\ }\href {\doibase
  10.1103/PhysRevD.87.115002} {\bibfield  {journal} {\bibinfo  {journal} {Phys.
  Rev. D}\ }\textbf {\bibinfo {volume} {87}},\ \bibinfo {pages} {115002}
  (\bibinfo {year} {2013})},\ \Eprint {http://arxiv.org/abs/1301.2336}
  {arXiv:1301.2336 [hep-ph]} \BibitemShut {NoStop}%
\bibitem [{\citenamefont {Fornal}\ and\ \citenamefont
  {Grinstein}(2017)}]{Fornal:2017xcj}%
  \BibitemOpen
  \bibfield  {author} {\bibinfo {author} {\bibfnamefont {B.}~\bibnamefont
  {Fornal}}\ and\ \bibinfo {author} {\bibfnamefont {B.}~\bibnamefont
  {Grinstein}},\ }\href {\doibase 10.1103/PhysRevLett.119.241801} {\bibfield
  {journal} {\bibinfo  {journal} {Phys. Rev. Lett.}\ }\textbf {\bibinfo
  {volume} {119}},\ \bibinfo {pages} {241801} (\bibinfo {year} {2017})},\
  \Eprint {http://arxiv.org/abs/1706.08535} {arXiv:1706.08535 [hep-ph]}
  \BibitemShut {NoStop}%
\bibitem [{\citenamefont {Cacciapaglia}\ \emph {et~al.}(2021)\citenamefont
  {Cacciapaglia}, \citenamefont {Cornell}, \citenamefont {Cot},\ and\
  \citenamefont {Deandrea}}]{Cacciapaglia:2020qky}%
  \BibitemOpen
  \bibfield  {author} {\bibinfo {author} {\bibfnamefont {G.}~\bibnamefont
  {Cacciapaglia}}, \bibinfo {author} {\bibfnamefont {A.~S.}\ \bibnamefont
  {Cornell}}, \bibinfo {author} {\bibfnamefont {C.}~\bibnamefont {Cot}}, \ and\
  \bibinfo {author} {\bibfnamefont {A.}~\bibnamefont {Deandrea}},\ }\href
  {\doibase 10.1103/PhysRevD.104.075012} {\bibfield  {journal} {\bibinfo
  {journal} {Phys. Rev. D}\ }\textbf {\bibinfo {volume} {104}},\ \bibinfo
  {pages} {075012} (\bibinfo {year} {2021})},\ \Eprint
  {http://arxiv.org/abs/2012.14732} {arXiv:2012.14732 [hep-th]} \BibitemShut
  {NoStop}%
\bibitem [{\citenamefont {Yanagida}(1995)}]{Yanagida:1994vq}%
  \BibitemOpen
  \bibfield  {author} {\bibinfo {author} {\bibfnamefont {T.}~\bibnamefont
  {Yanagida}},\ }\href {\doibase 10.1016/0370-2693(94)01500-C} {\bibfield
  {journal} {\bibinfo  {journal} {Phys. Lett. B}\ }\textbf {\bibinfo {volume}
  {344}},\ \bibinfo {pages} {211} (\bibinfo {year} {1995})},\ \Eprint
  {http://arxiv.org/abs/hep-ph/9409329} {arXiv:hep-ph/9409329} \BibitemShut
  {NoStop}%
\bibitem [{\citenamefont {Hotta}\ \emph {et~al.}(1996)\citenamefont {Hotta},
  \citenamefont {Izawa},\ and\ \citenamefont {Yanagida}}]{Hotta:1995cd}%
  \BibitemOpen
  \bibfield  {author} {\bibinfo {author} {\bibfnamefont {T.}~\bibnamefont
  {Hotta}}, \bibinfo {author} {\bibfnamefont {K.~I.}\ \bibnamefont {Izawa}}, \
  and\ \bibinfo {author} {\bibfnamefont {T.}~\bibnamefont {Yanagida}},\ }\href
  {\doibase 10.1103/PhysRevD.53.3913} {\bibfield  {journal} {\bibinfo
  {journal} {Phys. Rev. D}\ }\textbf {\bibinfo {volume} {53}},\ \bibinfo
  {pages} {3913} (\bibinfo {year} {1996})},\ \Eprint
  {http://arxiv.org/abs/hep-ph/9509201} {arXiv:hep-ph/9509201} \BibitemShut
  {NoStop}%
\bibitem [{\citenamefont {Ibe}\ and\ \citenamefont
  {Watari}(2003)}]{Ibe:2003ys}%
  \BibitemOpen
  \bibfield  {author} {\bibinfo {author} {\bibfnamefont {M.}~\bibnamefont
  {Ibe}}\ and\ \bibinfo {author} {\bibfnamefont {T.}~\bibnamefont {Watari}},\
  }\href {\doibase 10.1103/PhysRevD.67.114021} {\bibfield  {journal} {\bibinfo
  {journal} {Phys. Rev. D}\ }\textbf {\bibinfo {volume} {67}},\ \bibinfo
  {pages} {114021} (\bibinfo {year} {2003})},\ \Eprint
  {http://arxiv.org/abs/hep-ph/0303123} {arXiv:hep-ph/0303123} \BibitemShut
  {NoStop}%
\bibitem [{\citenamefont {'t~Hooft}(1976)}]{tHooft:1976rip}%
  \BibitemOpen
  \bibfield  {author} {\bibinfo {author} {\bibfnamefont {G.}~\bibnamefont
  {'t~Hooft}},\ }\href {\doibase 10.1103/PhysRevLett.37.8} {\bibfield
  {journal} {\bibinfo  {journal} {Phys. Rev. Lett.}\ }\textbf {\bibinfo
  {volume} {37}},\ \bibinfo {pages} {8} (\bibinfo {year} {1976})}\BibitemShut
  {NoStop}%
\bibitem [{\citenamefont {Hawking}(1987)}]{Hawking:1987mz}%
  \BibitemOpen
  \bibfield  {author} {\bibinfo {author} {\bibfnamefont {S.~W.}\ \bibnamefont
  {Hawking}},\ }\href {\doibase 10.1016/0370-2693(87)90028-1} {\bibfield
  {journal} {\bibinfo  {journal} {Phys. Lett. B}\ }\textbf {\bibinfo {volume}
  {195}},\ \bibinfo {pages} {337} (\bibinfo {year} {1987})}\BibitemShut
  {NoStop}%
\bibitem [{\citenamefont {Lavrelashvili}\ \emph {et~al.}(1987)\citenamefont
  {Lavrelashvili}, \citenamefont {Rubakov},\ and\ \citenamefont
  {Tinyakov}}]{Lavrelashvili:1987jg}%
  \BibitemOpen
  \bibfield  {author} {\bibinfo {author} {\bibfnamefont {G.~V.}\ \bibnamefont
  {Lavrelashvili}}, \bibinfo {author} {\bibfnamefont {V.~A.}\ \bibnamefont
  {Rubakov}}, \ and\ \bibinfo {author} {\bibfnamefont {P.~G.}\ \bibnamefont
  {Tinyakov}},\ }\href@noop {} {\bibfield  {journal} {\bibinfo  {journal} {JETP
  Lett.}\ }\textbf {\bibinfo {volume} {46}},\ \bibinfo {pages} {167} (\bibinfo
  {year} {1987})}\BibitemShut {NoStop}%
\bibitem [{\citenamefont {Giddings}\ and\ \citenamefont
  {Strominger}(1988)}]{Giddings:1988cx}%
  \BibitemOpen
  \bibfield  {author} {\bibinfo {author} {\bibfnamefont {S.~B.}\ \bibnamefont
  {Giddings}}\ and\ \bibinfo {author} {\bibfnamefont {A.}~\bibnamefont
  {Strominger}},\ }\href {\doibase 10.1016/0550-3213(88)90109-5} {\bibfield
  {journal} {\bibinfo  {journal} {Nucl. Phys. B}\ }\textbf {\bibinfo {volume}
  {307}},\ \bibinfo {pages} {854} (\bibinfo {year} {1988})}\BibitemShut
  {NoStop}%
\bibitem [{\citenamefont {Coleman}(1988)}]{Coleman:1988tj}%
  \BibitemOpen
  \bibfield  {author} {\bibinfo {author} {\bibfnamefont {S.~R.}\ \bibnamefont
  {Coleman}},\ }\href {\doibase 10.1016/0550-3213(88)90097-1} {\bibfield
  {journal} {\bibinfo  {journal} {Nucl. Phys. B}\ }\textbf {\bibinfo {volume}
  {310}},\ \bibinfo {pages} {643} (\bibinfo {year} {1988})}\BibitemShut
  {NoStop}%
\bibitem [{\citenamefont {Gilbert}(1989)}]{Gilbert:1989nq}%
  \BibitemOpen
  \bibfield  {author} {\bibinfo {author} {\bibfnamefont {G.}~\bibnamefont
  {Gilbert}},\ }\href {\doibase 10.1016/0550-3213(89)90097-7} {\bibfield
  {journal} {\bibinfo  {journal} {Nucl. Phys. B}\ }\textbf {\bibinfo {volume}
  {328}},\ \bibinfo {pages} {159} (\bibinfo {year} {1989})}\BibitemShut
  {NoStop}%
\bibitem [{\citenamefont {Kallosh}\ \emph {et~al.}(1995)\citenamefont
  {Kallosh}, \citenamefont {Linde}, \citenamefont {Linde},\ and\ \citenamefont
  {Susskind}}]{Kallosh:1995hi}%
  \BibitemOpen
  \bibfield  {author} {\bibinfo {author} {\bibfnamefont {R.}~\bibnamefont
  {Kallosh}}, \bibinfo {author} {\bibfnamefont {A.~D.}\ \bibnamefont {Linde}},
  \bibinfo {author} {\bibfnamefont {D.~A.}\ \bibnamefont {Linde}}, \ and\
  \bibinfo {author} {\bibfnamefont {L.}~\bibnamefont {Susskind}},\ }\href
  {\doibase 10.1103/PhysRevD.52.912} {\bibfield  {journal} {\bibinfo  {journal}
  {Phys. Rev. D}\ }\textbf {\bibinfo {volume} {52}},\ \bibinfo {pages} {912}
  (\bibinfo {year} {1995})},\ \Eprint {http://arxiv.org/abs/hep-th/9502069}
  {arXiv:hep-th/9502069} \BibitemShut {NoStop}%
\bibitem [{\citenamefont {Banks}\ and\ \citenamefont
  {Seiberg}(2011)}]{Banks:2010zn}%
  \BibitemOpen
  \bibfield  {author} {\bibinfo {author} {\bibfnamefont {T.}~\bibnamefont
  {Banks}}\ and\ \bibinfo {author} {\bibfnamefont {N.}~\bibnamefont
  {Seiberg}},\ }\href {\doibase 10.1103/PhysRevD.83.084019} {\bibfield
  {journal} {\bibinfo  {journal} {Phys. Rev. D}\ }\textbf {\bibinfo {volume}
  {83}},\ \bibinfo {pages} {084019} (\bibinfo {year} {2011})},\ \Eprint
  {http://arxiv.org/abs/1011.5120} {arXiv:1011.5120 [hep-th]} \BibitemShut
  {NoStop}%
\bibitem [{\citenamefont {Minkowski}(1977)}]{Minkowski:1977sc}%
  \BibitemOpen
  \bibfield  {author} {\bibinfo {author} {\bibfnamefont {P.}~\bibnamefont
  {Minkowski}},\ }\href {\doibase 10.1016/0370-2693(77)90435-X} {\bibfield
  {journal} {\bibinfo  {journal} {Phys. Lett. B}\ }\textbf {\bibinfo {volume}
  {67}},\ \bibinfo {pages} {421} (\bibinfo {year} {1977})}\BibitemShut
  {NoStop}%
\bibitem [{\citenamefont {Yanagida}(1979{\natexlab{a}})}]{Yanagida:1979as}%
  \BibitemOpen
  \bibfield  {author} {\bibinfo {author} {\bibfnamefont {T.}~\bibnamefont
  {Yanagida}},\ }\href@noop {} {\bibfield  {journal} {\bibinfo  {journal}
  {Conf. Proc. C}\ }\textbf {\bibinfo {volume} {7902131}},\ \bibinfo {pages}
  {95} (\bibinfo {year} {1979}{\natexlab{a}})}\BibitemShut {NoStop}%
\bibitem [{\citenamefont {Yanagida}(1979{\natexlab{b}})}]{Yanagida:1979gs}%
  \BibitemOpen
  \bibfield  {author} {\bibinfo {author} {\bibfnamefont {T.}~\bibnamefont
  {Yanagida}},\ }\href {\doibase 10.1103/PhysRevD.20.2986} {\bibfield
  {journal} {\bibinfo  {journal} {Phys. Rev. D}\ }\textbf {\bibinfo {volume}
  {20}},\ \bibinfo {pages} {2986} (\bibinfo {year}
  {1979}{\natexlab{b}})}\BibitemShut {NoStop}%
\bibitem [{\citenamefont {Gell-Mann}\ \emph {et~al.}(1979)\citenamefont
  {Gell-Mann}, \citenamefont {Ramond},\ and\ \citenamefont
  {Slansky}}]{Gell-Mann:1979vob}%
  \BibitemOpen
  \bibfield  {author} {\bibinfo {author} {\bibfnamefont {M.}~\bibnamefont
  {Gell-Mann}}, \bibinfo {author} {\bibfnamefont {P.}~\bibnamefont {Ramond}}, \
  and\ \bibinfo {author} {\bibfnamefont {R.}~\bibnamefont {Slansky}},\
  }\href@noop {} {\bibfield  {journal} {\bibinfo  {journal} {Conf. Proc. C}\
  }\textbf {\bibinfo {volume} {790927}},\ \bibinfo {pages} {315} (\bibinfo
  {year} {1979})},\ \Eprint {http://arxiv.org/abs/1306.4669} {arXiv:1306.4669
  [hep-th]} \BibitemShut {NoStop}%
\bibitem [{\citenamefont {Glashow}(1980)}]{Glashow:1979nm}%
  \BibitemOpen
  \bibfield  {author} {\bibinfo {author} {\bibfnamefont {S.~L.}\ \bibnamefont
  {Glashow}},\ }\href {\doibase 10.1007/978-1-4684-7197-7_15} {\bibfield
  {journal} {\bibinfo  {journal} {NATO Sci. Ser. B}\ }\textbf {\bibinfo
  {volume} {61}},\ \bibinfo {pages} {687} (\bibinfo {year} {1980})}\BibitemShut
  {NoStop}%
\bibitem [{\citenamefont {Mohapatra}\ and\ \citenamefont
  {Senjanovic}(1980)}]{Mohapatra:1979ia}%
  \BibitemOpen
  \bibfield  {author} {\bibinfo {author} {\bibfnamefont {R.~N.}\ \bibnamefont
  {Mohapatra}}\ and\ \bibinfo {author} {\bibfnamefont {G.}~\bibnamefont
  {Senjanovic}},\ }\href {\doibase 10.1103/PhysRevLett.44.912} {\bibfield
  {journal} {\bibinfo  {journal} {Phys. Rev. Lett.}\ }\textbf {\bibinfo
  {volume} {44}},\ \bibinfo {pages} {912} (\bibinfo {year} {1980})}\BibitemShut
  {NoStop}%
\bibitem [{\citenamefont {Nagata}\ and\ \citenamefont
  {Shirai}(2014)}]{Nagata:2013sba}%
  \BibitemOpen
  \bibfield  {author} {\bibinfo {author} {\bibfnamefont {N.}~\bibnamefont
  {Nagata}}\ and\ \bibinfo {author} {\bibfnamefont {S.}~\bibnamefont
  {Shirai}},\ }\href {\doibase 10.1007/JHEP03(2014)049} {\bibfield  {journal}
  {\bibinfo  {journal} {JHEP}\ }\textbf {\bibinfo {volume} {03}},\ \bibinfo
  {pages} {049} (\bibinfo {year} {2014})},\ \Eprint
  {http://arxiv.org/abs/1312.7854} {arXiv:1312.7854 [hep-ph]} \BibitemShut
  {NoStop}%
\bibitem [{\citenamefont {Aoki}\ \emph {et~al.}(2017)\citenamefont {Aoki},
  \citenamefont {Izubuchi}, \citenamefont {Shintani},\ and\ \citenamefont
  {Soni}}]{Aoki:2017puj}%
  \BibitemOpen
  \bibfield  {author} {\bibinfo {author} {\bibfnamefont {Y.}~\bibnamefont
  {Aoki}}, \bibinfo {author} {\bibfnamefont {T.}~\bibnamefont {Izubuchi}},
  \bibinfo {author} {\bibfnamefont {E.}~\bibnamefont {Shintani}}, \ and\
  \bibinfo {author} {\bibfnamefont {A.}~\bibnamefont {Soni}},\ }\href {\doibase
  10.1103/PhysRevD.96.014506} {\bibfield  {journal} {\bibinfo  {journal} {Phys.
  Rev. D}\ }\textbf {\bibinfo {volume} {96}},\ \bibinfo {pages} {014506}
  (\bibinfo {year} {2017})},\ \Eprint {http://arxiv.org/abs/1705.01338}
  {arXiv:1705.01338 [hep-lat]} \BibitemShut {NoStop}%
\bibitem [{\citenamefont {Yoo}\ \emph {et~al.}(2022)\citenamefont {Yoo},
  \citenamefont {Aoki}, \citenamefont {Boyle}, \citenamefont {Izubuchi},
  \citenamefont {Soni},\ and\ \citenamefont {Syritsyn}}]{Yoo:2021gql}%
  \BibitemOpen
  \bibfield  {author} {\bibinfo {author} {\bibfnamefont {J.-S.}\ \bibnamefont
  {Yoo}}, \bibinfo {author} {\bibfnamefont {Y.}~\bibnamefont {Aoki}}, \bibinfo
  {author} {\bibfnamefont {P.}~\bibnamefont {Boyle}}, \bibinfo {author}
  {\bibfnamefont {T.}~\bibnamefont {Izubuchi}}, \bibinfo {author}
  {\bibfnamefont {A.}~\bibnamefont {Soni}}, \ and\ \bibinfo {author}
  {\bibfnamefont {S.}~\bibnamefont {Syritsyn}},\ }\href {\doibase
  10.1103/PhysRevD.105.074501} {\bibfield  {journal} {\bibinfo  {journal}
  {Phys. Rev. D}\ }\textbf {\bibinfo {volume} {105}},\ \bibinfo {pages}
  {074501} (\bibinfo {year} {2022})},\ \Eprint
  {http://arxiv.org/abs/2111.01608} {arXiv:2111.01608 [hep-lat]} \BibitemShut
  {NoStop}%
\bibitem [{\citenamefont {Kobayashi}\ \emph {et~al.}(2005)\citenamefont
  {Kobayashi} \emph {et~al.}}]{Super-Kamiokande:2005lev}%
  \BibitemOpen
  \bibfield  {author} {\bibinfo {author} {\bibfnamefont {K.}~\bibnamefont
  {Kobayashi}} \emph {et~al.} (\bibinfo {collaboration} {Super-Kamiokande}),\
  }\href {\doibase 10.1103/PhysRevD.72.052007} {\bibfield  {journal} {\bibinfo
  {journal} {Phys. Rev. D}\ }\textbf {\bibinfo {volume} {72}},\ \bibinfo
  {pages} {052007} (\bibinfo {year} {2005})},\ \Eprint
  {http://arxiv.org/abs/hep-ex/0502026} {arXiv:hep-ex/0502026} \BibitemShut
  {NoStop}%
\bibitem [{\citenamefont {Regis}\ \emph {et~al.}(2012)\citenamefont {Regis}
  \emph {et~al.}}]{Super-Kamiokande:2012zik}%
  \BibitemOpen
  \bibfield  {author} {\bibinfo {author} {\bibfnamefont {C.}~\bibnamefont
  {Regis}} \emph {et~al.} (\bibinfo {collaboration} {Super-Kamiokande}),\
  }\href {\doibase 10.1103/PhysRevD.86.012006} {\bibfield  {journal} {\bibinfo
  {journal} {Phys. Rev. D}\ }\textbf {\bibinfo {volume} {86}},\ \bibinfo
  {pages} {012006} (\bibinfo {year} {2012})},\ \Eprint
  {http://arxiv.org/abs/1205.6538} {arXiv:1205.6538 [hep-ex]} \BibitemShut
  {NoStop}%
\bibitem [{\citenamefont {Abe}\ \emph {et~al.}(2014{\natexlab{a}})\citenamefont
  {Abe} \emph {et~al.}}]{Super-Kamiokande:2013rwg}%
  \BibitemOpen
  \bibfield  {author} {\bibinfo {author} {\bibfnamefont {K.}~\bibnamefont
  {Abe}} \emph {et~al.} (\bibinfo {collaboration} {Super-Kamiokande}),\ }\href
  {\doibase 10.1103/PhysRevLett.113.121802} {\bibfield  {journal} {\bibinfo
  {journal} {Phys. Rev. Lett.}\ }\textbf {\bibinfo {volume} {113}},\ \bibinfo
  {pages} {121802} (\bibinfo {year} {2014}{\natexlab{a}})},\ \Eprint
  {http://arxiv.org/abs/1305.4391} {arXiv:1305.4391 [hep-ex]} \BibitemShut
  {NoStop}%
\bibitem [{\citenamefont {Abe}\ \emph {et~al.}(2014{\natexlab{b}})\citenamefont
  {Abe} \emph {et~al.}}]{Super-Kamiokande:2014otb}%
  \BibitemOpen
  \bibfield  {author} {\bibinfo {author} {\bibfnamefont {K.}~\bibnamefont
  {Abe}} \emph {et~al.} (\bibinfo {collaboration} {Super-Kamiokande}),\ }\href
  {\doibase 10.1103/PhysRevD.90.072005} {\bibfield  {journal} {\bibinfo
  {journal} {Phys. Rev. D}\ }\textbf {\bibinfo {volume} {90}},\ \bibinfo
  {pages} {072005} (\bibinfo {year} {2014}{\natexlab{b}})},\ \Eprint
  {http://arxiv.org/abs/1408.1195} {arXiv:1408.1195 [hep-ex]} \BibitemShut
  {NoStop}%
\bibitem [{\citenamefont {Abe}\ \emph {et~al.}(2017)\citenamefont {Abe} \emph
  {et~al.}}]{Super-Kamiokande:2017gev}%
  \BibitemOpen
  \bibfield  {author} {\bibinfo {author} {\bibfnamefont {K.}~\bibnamefont
  {Abe}} \emph {et~al.} (\bibinfo {collaboration} {Super-Kamiokande}),\ }\href
  {\doibase 10.1103/PhysRevD.96.012003} {\bibfield  {journal} {\bibinfo
  {journal} {Phys. Rev. D}\ }\textbf {\bibinfo {volume} {96}},\ \bibinfo
  {pages} {012003} (\bibinfo {year} {2017})},\ \Eprint
  {http://arxiv.org/abs/1705.07221} {arXiv:1705.07221 [hep-ex]} \BibitemShut
  {NoStop}%
\bibitem [{\citenamefont {Hall}\ \emph {et~al.}(2014)\citenamefont {Hall},
  \citenamefont {Nomura},\ and\ \citenamefont {Shirai}}]{Hall:2014vga}%
  \BibitemOpen
  \bibfield  {author} {\bibinfo {author} {\bibfnamefont {L.~J.}\ \bibnamefont
  {Hall}}, \bibinfo {author} {\bibfnamefont {Y.}~\bibnamefont {Nomura}}, \ and\
  \bibinfo {author} {\bibfnamefont {S.}~\bibnamefont {Shirai}},\ }\href
  {\doibase 10.1007/JHEP06(2014)137} {\bibfield  {journal} {\bibinfo  {journal}
  {JHEP}\ }\textbf {\bibinfo {volume} {06}},\ \bibinfo {pages} {137} (\bibinfo
  {year} {2014})},\ \Eprint {http://arxiv.org/abs/1403.8138} {arXiv:1403.8138
  [hep-ph]} \BibitemShut {NoStop}%
\bibitem [{\citenamefont {Rubakov}(1982)}]{Rubakov:1982fp}%
  \BibitemOpen
  \bibfield  {author} {\bibinfo {author} {\bibfnamefont {V.~A.}\ \bibnamefont
  {Rubakov}},\ }\href {\doibase 10.1016/0550-3213(82)90034-7} {\bibfield
  {journal} {\bibinfo  {journal} {Nucl. Phys. B}\ }\textbf {\bibinfo {volume}
  {203}},\ \bibinfo {pages} {311} (\bibinfo {year} {1982})}\BibitemShut
  {NoStop}%
\bibitem [{\citenamefont {Callan}(1982)}]{Callan:1982ah}%
  \BibitemOpen
  \bibfield  {author} {\bibinfo {author} {\bibfnamefont {C.~G.}\ \bibnamefont
  {Callan}, \bibfnamefont {Jr.}},\ }\href {\doibase 10.1103/PhysRevD.25.2141}
  {\bibfield  {journal} {\bibinfo  {journal} {Phys. Rev. D}\ }\textbf {\bibinfo
  {volume} {25}},\ \bibinfo {pages} {2141} (\bibinfo {year}
  {1982})}\BibitemShut {NoStop}%
\bibitem [{\citenamefont {Goodman}\ and\ \citenamefont
  {Witten}(1986)}]{GoodMan:1985bw}%
  \BibitemOpen
  \bibfield  {author} {\bibinfo {author} {\bibfnamefont {M.~W.}\ \bibnamefont
  {Goodman}}\ and\ \bibinfo {author} {\bibfnamefont {E.}~\bibnamefont
  {Witten}},\ }\href {\doibase 10.1016/0550-3213(86)90352-4} {\bibfield
  {journal} {\bibinfo  {journal} {Nucl. Phys. B}\ }\textbf {\bibinfo {volume}
  {271}},\ \bibinfo {pages} {21} (\bibinfo {year} {1986})}\BibitemShut
  {NoStop}%
\bibitem [{\citenamefont {Sartore}\ and\ \citenamefont
  {Schienbein}(2021)}]{Sartore:2020gou}%
  \BibitemOpen
  \bibfield  {author} {\bibinfo {author} {\bibfnamefont {L.}~\bibnamefont
  {Sartore}}\ and\ \bibinfo {author} {\bibfnamefont {I.}~\bibnamefont
  {Schienbein}},\ }\href {\doibase 10.1016/j.cpc.2020.107819} {\bibfield
  {journal} {\bibinfo  {journal} {Comput. Phys. Commun.}\ }\textbf {\bibinfo
  {volume} {261}},\ \bibinfo {pages} {107819} (\bibinfo {year} {2021})},\
  \Eprint {http://arxiv.org/abs/2007.12700} {arXiv:2007.12700 [hep-ph]}
  \BibitemShut {NoStop}%
\bibitem [{\citenamefont {Ibanez}\ and\ \citenamefont
  {Ross}(1991)}]{Ibanez:1991hv}%
  \BibitemOpen
  \bibfield  {author} {\bibinfo {author} {\bibfnamefont {L.~E.}\ \bibnamefont
  {Ibanez}}\ and\ \bibinfo {author} {\bibfnamefont {G.~G.}\ \bibnamefont
  {Ross}},\ }\href {\doibase 10.1016/0370-2693(91)91614-2} {\bibfield
  {journal} {\bibinfo  {journal} {Phys. Lett. B}\ }\textbf {\bibinfo {volume}
  {260}},\ \bibinfo {pages} {291} (\bibinfo {year} {1991})}\BibitemShut
  {NoStop}%
\bibitem [{\citenamefont {Ibanez}\ and\ \citenamefont
  {Ross}(1992)}]{Ibanez:1991pr}%
  \BibitemOpen
  \bibfield  {author} {\bibinfo {author} {\bibfnamefont {L.~E.}\ \bibnamefont
  {Ibanez}}\ and\ \bibinfo {author} {\bibfnamefont {G.~G.}\ \bibnamefont
  {Ross}},\ }\href {\doibase 10.1016/0550-3213(92)90195-H} {\bibfield
  {journal} {\bibinfo  {journal} {Nucl. Phys. B}\ }\textbf {\bibinfo {volume}
  {368}},\ \bibinfo {pages} {3} (\bibinfo {year} {1992})}\BibitemShut {NoStop}%
\bibitem [{\citenamefont {Ibanez}(1993)}]{Ibanez:1992ji}%
  \BibitemOpen
  \bibfield  {author} {\bibinfo {author} {\bibfnamefont {L.~E.}\ \bibnamefont
  {Ibanez}},\ }\href {\doibase 10.1016/0550-3213(93)90111-2} {\bibfield
  {journal} {\bibinfo  {journal} {Nucl. Phys. B}\ }\textbf {\bibinfo {volume}
  {398}},\ \bibinfo {pages} {301} (\bibinfo {year} {1993})},\ \Eprint
  {http://arxiv.org/abs/hep-ph/9210211} {arXiv:hep-ph/9210211} \BibitemShut
  {NoStop}%
\bibitem [{\citenamefont {Csaki}\ and\ \citenamefont
  {Murayama}(1998)}]{Csaki:1997aw}%
  \BibitemOpen
  \bibfield  {author} {\bibinfo {author} {\bibfnamefont {C.}~\bibnamefont
  {Csaki}}\ and\ \bibinfo {author} {\bibfnamefont {H.}~\bibnamefont
  {Murayama}},\ }\href {\doibase 10.1016/S0550-3213(97)00839-0} {\bibfield
  {journal} {\bibinfo  {journal} {Nucl. Phys. B}\ }\textbf {\bibinfo {volume}
  {515}},\ \bibinfo {pages} {114} (\bibinfo {year} {1998})},\ \Eprint
  {http://arxiv.org/abs/hep-th/9710105} {arXiv:hep-th/9710105} \BibitemShut
  {NoStop}%
\bibitem [{\citenamefont {Banks}\ and\ \citenamefont
  {Dine}(1992)}]{Banks:1991xj}%
  \BibitemOpen
  \bibfield  {author} {\bibinfo {author} {\bibfnamefont {T.}~\bibnamefont
  {Banks}}\ and\ \bibinfo {author} {\bibfnamefont {M.}~\bibnamefont {Dine}},\
  }\href {\doibase 10.1103/PhysRevD.45.1424} {\bibfield  {journal} {\bibinfo
  {journal} {Phys. Rev. D}\ }\textbf {\bibinfo {volume} {45}},\ \bibinfo
  {pages} {1424} (\bibinfo {year} {1992})},\ \Eprint
  {http://arxiv.org/abs/hep-th/9109045} {arXiv:hep-th/9109045} \BibitemShut
  {NoStop}%
\end{thebibliography}%

\end{document}